\documentclass[prx,aps,showpacs,twocolumn,preprintnumbers,amsmath,amssymb,superscriptaddress,longbibliography]{revtex4-2}

\usepackage{CJKutf8}
\usepackage[colorlinks,citecolor=blue,urlcolor=blue,linkcolor=blue]{hyperref}
\usepackage[all]{hypcap} 

\usepackage{tikzit}

\tikzstyle{hadamard}=[fill=yellow, draw, inner sep=0.6mm, minimum height=1.5mm, minimum width=1.5mm, shape=rectangle, tikzit shape=rectangle, tikzit category=ZH-pf, line width=0.5pt]
\tikzstyle{dd hadamard}=[hadamard, line width=1.6pt, fill=yellow, tikzit shape=rectangle, tikzit fill={rgb,255: red,222; green,222; blue,0}, tikzit draw=black]
\tikzstyle{Z dot}=[inner sep=0mm, minimum size=2mm, shape=circle, draw=black, fill={zx_green}, tikzit fill=green, tikzit draw=white]
\tikzstyle{Z phase dot}=[minimum size=5mm, font={\footnotesize\boldmath}, shape=rectangle, rounded corners=2mm, inner sep=0.2mm, outer sep=-2mm, scale=0.8, tikzit shape=circle, draw=black, fill={zx_green}, tikzit draw=blue, tikzit fill=green]
\tikzstyle{dd Z}=[Z dot, line width=1.6pt, tikzit draw=white, tikzit fill={rgb,255: red,26; green,182; blue,23}]
\tikzstyle{dd Z phase}=[Z phase dot, line width=1.6pt, tikzit draw=black, tikzit fill={rgb,255: red,26; green,182; blue,23}]
\tikzstyle{X dot}=[Z dot, shape=circle, draw=black, fill={zx_red}, tikzit fill=red, tikzit draw=white]
\tikzstyle{X phase dot}=[Z phase dot, tikzit shape=circle, tikzit draw=blue, fill={zx_red}, font={\footnotesize\color{black}\boldmath}, tikzit fill=red]
\tikzstyle{dd X}=[X dot, line width=1.6pt, tikzit draw=white, tikzit fill={rgb,255: red,200; green,0; blue,0}]
\tikzstyle{dd X phase}=[X phase dot, line width=1.6pt, tikzit draw=black, tikzit fill={rgb,255: red,200; green,0; blue,0}]
\tikzstyle{box}=[shape=rectangle, text height=1.5ex, text depth=0.25ex, yshift=0.2mm, fill=white, draw=black, minimum height=3mm, minimum width=5mm, font={\small}, line width=0.5pt]
\tikzstyle{gray box}=[shape=rectangle, text height=1.5ex, text depth=0.25ex, yshift=0.2mm, fill=lightgray, draw=black, minimum height=3mm, minimum width=5mm, font={\small}, line width=0.5pt]
\tikzstyle{orange_box}=[shape=rectangle, text height=1.5ex, text depth=0.25ex, yshift=0.2mm, fill=orange, draw=black, minimum height=3mm, minimum width=5mm, font={\small}, line width=0.5pt]
\tikzstyle{very tall box}=[box, minimum height=2cm, minimum width=8mm, shape=rectangle, tikzit fill=white, tikzit draw=black, line width=0.5pt]
\tikzstyle{tall box}=[box, minimum height=1.5cm, minimum width=8mm, shape=rectangle, tikzit fill=white, tikzit draw=black, line width=0.5pt]
\tikzstyle{mid box}=[box, minimum height=0.8cm, minimum width=5mm, shape=rectangle, tikzit fill=white, tikzit draw=black, line width=0.5pt]
\tikzstyle{dd box}=[box, line width=1.6pt, tikzit fill=black, tikzit draw=black, shape=rectangle]
\tikzstyle{dd tall box}=[box, minimum height=1.5cm, minimum width=8mm, line width=1.6pt, tikzit fill=black, tikzit draw=black, shape=rectangle]
\tikzstyle{dot}=[inner sep=0.3mm, minimum width=2mm, minimum height=2mm, draw, shape=circle, font={\footnotesize}, tikzit fill=magenta]
\tikzstyle{blue large dot}=[minimum size=5mm, font={\footnotesize\boldmath}, shape=rectangle, rounded corners=2mm, inner sep=0.2mm, outer sep=-2mm, scale=0.8, tikzit shape=circle, draw=black, fill={rgb,255: red,102; green,178; blue,255}, tikzit draw=blue, tikzit fill=blue]
\tikzstyle{red large dot}=[minimum size=5mm, font={\footnotesize\boldmath}, shape=rectangle, rounded corners=2mm, inner sep=0.2mm, outer sep=-2mm, scale=0.8, tikzit shape=circle, draw=black, fill={rgb,255: red,255; green,153; blue,153}, tikzit draw=red, tikzit fill=red]
\tikzstyle{white dot}=[dot, fill=white, text depth=-0.2mm, tikzit category=ZH-pf, draw=black]
\tikzstyle{white phase dot}=[minimum size=5mm, font={\footnotesize\boldmath}, shape=rectangle, rounded corners=2mm, inner sep=0.2mm, outer sep=-2mm, scale=0.8, tikzit shape=circle, draw=black, fill=white, tikzit category=ZH-pf, tikzit fill=white, tikzit draw=blue]
\tikzstyle{gray dot}=[dot, fill={rgb,255: red,180; green,180; blue,180}, text depth=-0.2mm, tikzit category=ZH-pf]
\tikzstyle{gray phase dot}=[white phase dot, tikzit shape=circle, tikzit draw=blue, fill={rgb,255: red,180; green,180; blue,180}, font={\footnotesize\boldmath}]
\tikzstyle{black dot}=[dot, fill=black, text depth=-0.2mm, tikzit category=ZH-pf, draw=black]
\tikzstyle{black phase dot}=[minimum size=5mm, font={\footnotesize\boldmath}, shape=rectangle, rounded corners=2mm, inner sep=0.2mm, outer sep=-2mm, scale=0.8, tikzit shape=circle, draw=black, fill=black, tikzit category=ZH-pf, tikzit fill=black, tikzit draw=blue]
\tikzstyle{halfscalar}=[star, fill=black, draw=black, minimum size=8pt, inner sep=0pt]
\tikzstyle{H box}=[hadamard]
\tikzstyle{st}=[star, star points=5, fill=white, draw=black, inner sep=1.2pt, line width=1.2pt, tikzit fill=blue, tikzit draw=red, tikzit category=ZH-pf]
\tikzstyle{triangle}=[regular polygon, regular polygon sides=3, fill=white, draw=black, inner sep=0pt, minimum width=1em, tikzit draw=blue, tikzit category=ZH-pf, tikzit fill=cyan]
\tikzstyle{not}=[fill={rgb,255: red,180; green,180; blue,180}, draw=black, shape=circle, font={$\neg$}, dot]
\tikzstyle{vertex}=[inner sep=0mm, minimum size=1mm, shape=circle, draw=black, fill=black]
\tikzstyle{vertex set}=[inner sep=0mm, minimum size=1mm, shape=circle, draw=black, fill=white, font={\footnotesize\boldmath}]
\tikzstyle{wide point}=[fill=white, draw, shape=isosceles triangle, shape border rotate=-90, isosceles triangle stretches=true, inner sep=0pt, minimum width=1.5cm, minimum height=6.12mm, yshift=-0.0mm]
\tikzstyle{medium gray box}=[semilarge box, fill={rgb,255: red,180; green,180; blue,180}]
\tikzstyle{small box}=[rectangle, inline text, fill=white, draw, minimum height=5mm, yshift=-0.5mm, minimum width=5mm, font={\small}]
\tikzstyle{small gray box}=[small box, fill={rgb,255: red,180; green,180; blue,180}]
\tikzstyle{medium box}=[rectangle, inline text, fill=white, draw, minimum height=5mm, yshift=-0.5mm, minimum width=8mm, font={\small}]
\tikzstyle{ddot}=[line width=1.6pt, inner sep=0mm, minimum width=2.5mm, minimum height=2.5mm, draw, shape=circle]
\tikzstyle{dd white}=[ddot, fill=white, tikzit draw=green]
\tikzstyle{dd white phase}=[white phase dot, line width=1.6pt, tikzit draw=yellow]
\tikzstyle{dd gray}=[ddot, fill={rgb,255: red,180; green,180; blue,180}, tikzit draw=green]
\tikzstyle{dd gray phase}=[gray phase dot, line width=1.6pt, tikzit draw=yellow]

\tikzstyle{simple}=[-, line width=0.5pt, fill=none, draw=black, tikzit draw={rgb,255: red,162; green,162; blue,162}]
\tikzstyle{hadamard edge}=[-, dashed, dash pattern=on 2pt off 1pt, thick, draw=gray]
\tikzstyle{gray}=[-, draw={blue!60!white}, tikzit draw=blue]
\tikzstyle{blue}=[-, draw={blue!60!white}, tikzit draw=blue]
\tikzstyle{blue dashed}=[-, dashed, draw={blue!60!white}, tikzit draw=blue]
\tikzstyle{blue dashed directed}=[->, dashed, draw={blue!60!white}, tikzit draw=blue]
\tikzstyle{red}=[-, draw={red!60!white}, tikzit draw=red]
\tikzstyle{red directed}=[->, draw={red!60!white}, tikzit draw=red]
\tikzstyle{red thick}=[-, thick, draw={rgb,255: red,255; green,68; blue,68}]
\tikzstyle{red dashed}=[-, dashed, draw={red!60!white}, tikzit draw=red]
\tikzstyle{brace edge}=[-, tikzit draw=blue, decorate, decoration={brace,amplitude=1mm,raise=-1mm}]
\tikzstyle{diredge}=[->]
\tikzstyle{dashed diredge}=[->, dashed]
\tikzstyle{not edge}=[-, dashed, dash pattern=on 2pt off 1.5pt, thick, draw={rgb,255: red,255; green,68; blue,68}]
\tikzstyle{not diredge}=[->, dashed, dash pattern=on 2pt off 1.5pt, thick, draw={rgb,255: red,255; green,68; blue,68}]
\tikzstyle{double edge}=[-, double, shorten <=-1mm, shorten >=-1mm, double distance=2pt]
\tikzstyle{boldedge}=[-, line width=1.6pt, shorten <=-0.17mm, shorten >=-0.17mm, tikzit draw=blue]
\tikzstyle{dots}=[-, dashed, dash pattern=on 1.5pt off 1.0pt, tikzit draw=black, thick]
\tikzstyle{simple0.5}=[-, line width=0.5pt, fill=none, draw=black, tikzit draw={rgb,255: red,162; green,162; blue,162}]
\tikzstyle{simple0.6}=[-, line width=0.6pt, fill=none, draw=black, tikzit draw={rgb,255: red,162; green,162; blue,162}]
\tikzstyle{simple0.7}=[-, line width=0.7pt, fill=none, draw=black, tikzit draw={rgb,255: red,162; green,162; blue,162}]
\tikzstyle{simple0.8}=[-, line width=0.8pt, fill=none, draw=black, tikzit draw={rgb,255: red,162; green,162; blue,162}]
\tikzstyle{simple0.9}=[-, line width=0.9pt, fill=none, draw=black, tikzit draw={rgb,255: red,162; green,162; blue,162}]
\tikzstyle{simple1.0}=[-, line width=1.0pt, fill=none, draw=black, tikzit draw={rgb,255: red,162; green,162; blue,162}]
\tikzstyle{simple1.1}=[-, line width=1.1pt, fill=none, draw=black, tikzit draw={rgb,255: red,162; green,162; blue,162}]
\tikzstyle{simple1.2}=[-, line width=1.2pt, fill=none, draw=black, tikzit draw={rgb,255: red,162; green,162; blue,162}]
\tikzstyle{simple1.3}=[-, line width=1.3pt, fill=none, draw=black, tikzit draw={rgb,255: red,162; green,162; blue,162}]
\tikzstyle{simple1.4}=[-, line width=1.4pt, fill=none, draw=black, tikzit draw={rgb,255: red,162; green,162; blue,162}]
\tikzstyle{simple1.5}=[-, line width=1.5pt, fill=none, draw=black, tikzit draw={rgb,255: red,162; green,162; blue,162}]
\tikzstyle{simple1.6}=[-, line width=1.6pt, fill=none, draw=black, tikzit draw={rgb,255: red,162; green,162; blue,162}]
\tikzstyle{simple1.7}=[-, line width=1.7pt, fill=none, draw=black, tikzit draw={rgb,255: red,162; green,162; blue,162}]
\tikzstyle{simple1.8}=[-, line width=1.8pt, fill=none, draw=black, tikzit draw={rgb,255: red,162; green,162; blue,162}]
\tikzstyle{simple1.9}=[-, line width=1.9pt, fill=none, draw=black, tikzit draw={rgb,255: red,162; green,162; blue,162}]
\tikzstyle{simple2.0}=[-, line width=2.0pt, fill=none, draw=black, tikzit draw={rgb,255: red,162; green,162; blue,162}]
\tikzstyle{simple2.1}=[-, line width=2.1pt, fill=none, draw=black, tikzit draw={rgb,255: red,162; green,162; blue,162}]
\tikzstyle{simple2.2}=[-, line width=2.2pt, fill=none, draw=black, tikzit draw={rgb,255: red,162; green,162; blue,162}]
\tikzstyle{simple2.3}=[-, line width=2.3pt, fill=none, draw=black, tikzit draw={rgb,255: red,162; green,162; blue,162}]
\tikzstyle{simple2.4}=[-, line width=2.4pt, fill=none, draw=black, tikzit draw={rgb,255: red,162; green,162; blue,162}]
\tikzstyle{simple2.5}=[-, line width=2.5pt, fill=none, draw=black, tikzit draw={rgb,255: red,162; green,162; blue,162}]
\tikzstyle{simple2.6}=[-, line width=2.6pt, fill=none, draw=black, tikzit draw={rgb,255: red,162; green,162; blue,162}]
\tikzstyle{simple2.7}=[-, line width=2.7pt, fill=none, draw=black, tikzit draw={rgb,255: red,162; green,162; blue,162}]
\tikzstyle{simple2.8}=[-, line width=2.8pt, fill=none, draw=black, tikzit draw={rgb,255: red,162; green,162; blue,162}]
\tikzstyle{simple2.9}=[-, line width=2.9pt, fill=none, draw=black, tikzit draw={rgb,255: red,162; green,162; blue,162}]
\tikzstyle{simple3.0}=[-, line width=3.0pt, fill=none, draw=black, tikzit draw={rgb,255: red,162; green,162; blue,162}]
\tikzstyle{simple4.0}=[-, line width=4.0pt, fill=none, draw=black, tikzit draw={rgb,255: red,162; green,162; blue,162}]
\tikzstyle{simple8.0}=[-, line width=8.0pt, fill=none, draw=black, tikzit draw={rgb,255: red,162; green,162; blue,162}]
\input{zh.tikzdefs}

\usepackage{pifont}
\usepackage{graphicx}
\usepackage{bm}
\usepackage{dcolumn}
\usepackage{mathtools}
\usepackage{float}
\usepackage{physics}
\usepackage[utf8]{inputenc}
\usepackage[twocolumn,textwidth=510pt,columnsep=18pt,tmargin=1in,bmargin=1in]{geometry}
\usepackage[title]{appendix}
\usepackage{MnSymbol}

\newtheorem{theorem}{Theorem}
\newtheorem{definition}{Definition}
\newtheorem{corollary}{Corollary}
  
\definecolor{eggplant}{RGB}{180,33,147}

\begin{document}

\author{\begin{CJK*}{UTF8}{gbsn}Jielun Chen (陈捷伦)\end{CJK*}}
\email{jchen9@caltech.edu}
\affiliation{Department of Physics and Astronomy, University of California, Irvine, CA 92697-4575, USA}
\affiliation{Department of Physics, California Institute of Technology, Pasadena, CA 91125, USA}

\author{E.M.\ Stoudenmire}
\affiliation{Center for Computational Quantum Physics, Flatiron Institute, 162 5th Avenue, New York, NY 10010, USA}

\author{Steven R. White}
\affiliation{Department of Physics and Astronomy, University of California, Irvine, CA 92697-4575, USA}

\title{Quantum Fourier Transform Has Small Entanglement}
\begin{abstract}
    The Quantum Fourier Transform (QFT) is a key component of many important quantum algorithms, most famously as being the essential ingredient in Shor's algorithm for factoring products of primes. Given its remarkable capability, one would think it can introduce large entanglement to qubit systems and would be difficult to simulate classically. While early results showed QFT indeed has maximal operator entanglement, we show that this is entirely due to the bit reversal in the QFT. The core part of the QFT has Schmidt coefficients decaying exponentially quickly, and thus it can only generate a constant amount of entanglement regardless of the number of qubits. In addition, we show the entangling power of the QFT is the same as the time evolution of a Hamiltonian with exponentially decaying interactions, and thus a variant of the area law for dynamics can be used to understand the low entanglement intuitively. Using the low entanglement property of the QFT, we show that classical simulations of the QFT on a matrix product state with low bond dimension only take time linear in the number of qubits, providing a potential speedup over the classical fast Fourier transform (FFT) on many classes of functions. We demonstrate this speedup in test calculations on some simple functions. For data vectors of length $10^6$ to $10^8$, the speedup can be a few orders of magnitude.
\end{abstract}

\maketitle

\section{Introduction}
\label{section:intro}
The quantum Fourier transform (QFT) is one of the most important building blocks of quantum algorithms. Its power stems from the ability to find periodic structures of quantum states, which many quantum algorithms rely on to gain a provable advantage over classical algorithms. The most famous example is Shor's algorithm for factoring products of primes, where the QFT is used to find the period of the modular exponentiation function \cite{Shor_1997}. The QFT is also used in hidden subgroup problems \cite{Kitaev_1996, Hoyer_1999}, phase estimation \cite{Kitaev_1996}, simulation of quantum chemistry \cite{Kassal_2011}, solving linear systems of equations \cite{Harrow_2009}, quantum arithmetic \cite{Ruiz_Perez_2017}, and many other applications. Given the remarkable capability of the QFT, it would be natural to assume that it can generate large entanglement. Indeed, the exact operator Schmidt decomposition of the QFT has been found \cite{Nielsen_2003, Tyson_2003}, where the singular values are uniform, corresponding to maximal operator entanglement. However, it was noticed in numerical calculations in \cite{Woolfe_2014} that this maximal entanglement is entirely due to the bit reversal in the QFT. The core of the QFT has Schmidt coefficients which were found to decay very quickly, but the underlying reasons for this low entanglement were not studied, nor were any bounds on the decay proved. Since a reordering of the qubits can be kept track of very easily, whether on a quantum computer or in a classical simulation, the low entanglement of the core part of the QFT is very useful, for example allowing classical matrix product state (MPS) simulations to be used efficiently.  

In this paper, we develop an understanding of the low entanglement of the QFT and prove bounds on the Schmidt coefficients. We find that the low entanglement stems from the mapping of the Schmidt coefficients to the solution of the spectral concentration problem, a well-known problem in signal processing \cite{Slepian_1961, Landau_1961, Landau_1962, Slepian_1964, Slepian_1978} which aims to find a vector of a given length whose Fourier transform is maximally localized on a given frequency interval. By choosing a number of qubits $n$, one is discretizing the \emph{spectral-concentration} problem on an exponentially fine grid of $2^n$ points. One can also view the discretized problem as finding the singular values of sub-matrices of the discrete Fourier transform matrix. The rapid decay of the eigenvalues in spectral concentration is well-understood \cite{Slepian_1978, Karnik_2019_1, Matthysen_2016, Karnik_2019_2, Zhu_2015, Boulsane_2020, Karnik_2021}. 

Employing such previous results, we prove bounds on the Schmidt coefficients of the core part of the QFT, implying the QFT's low-entanglement structure. This allows us to construct an efficient matrix product operator (MPO) representation. The dimension of the MPO is not only nearly constant in the number of qubits, it is also remarkably small: an MPO of dimension 8 is sufficient for double precision accuracy. Therefore, by simply applying the QFT-MPO to an MPS with a low bond dimension, one can simulate the QFT in a time linear in the number of qubits, i.e. logarithmic in the Hilbert space size, and with a very small coefficient for the dominant term. We note our scheme is fundamentally different from previous efficient tensor network simulations of the QFT \cite{Aharonov_2006, Yoran_2007} in three ways: firstly, the low-rank structure of the QFT was not known and thus not used; secondly, the cited work considered the operator being applied and postselected on some states, while we are simulating the compression of the QFT operator itself; thirdly, the cited work considered the approximate QFT (AQFT) while we are simulating the full circuit of the QFT.

Our paper is organized as follows. In Section~\ref{subsection:QFT_LE} we present our results on the bounds of Schmidt coefficients of the QFT, and prove this bound in Section~\ref{subsection:QFT_LE_proof} by connecting to the spectral concentration problem. Then we provide a more intuitive explanation of the QFT's low entanglement in Section~\ref{subsection:area_law} by connecting to the area law for dynamics. In Section~\ref{subsection:TN_MPO} we briefly introduce the concept of tensor networks and the MPOs. Followed by that, in Section~\ref{subsection:QFT_TN} we introduce a tensor network representation of the QFT and an efficient algorithm for generating the MPO for the QFT. Finally, in Section~\ref{section:QFT_vs_FFT} we benchmark the QFT-MPO method against the fast Fourier transform (FFT) algorithm on classical computers, and demonstrate its benefits over FFT for many classes of functions. 

\section{Exponential Decay of the QFT's Schmidt Coefficients}
\label{section:QFT_schmidt}
In this section, we present our main theorem, the exponential decay of the Schmidt coefficients of the QFT with the output bits' order reversed. Then we prove the exponential decay by connecting to the spectral concentration problem. We also provide a more intuitive explanation of this low entanglement property by connecting it to the exponential decay of interactions in a Hamiltonian and the area law for dynamics.

\subsection{The Main Theorem}
\label{subsection:QFT_LE}
\begin{figure}[tb]
    \centering
    \includegraphics[width=1\columnwidth]{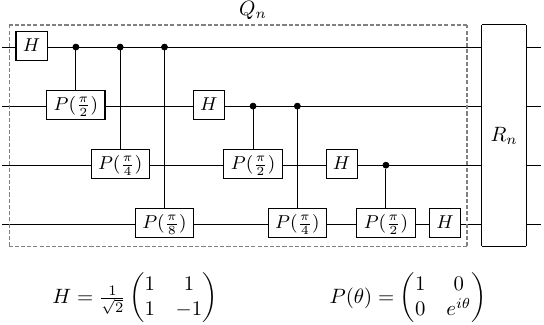}
    \caption{A 4-qubit example of the QFT circuit, i.e. $F_n$ with \mbox{$n=4$}, consisting of two parts, $Q_n$ and the bit reversal operation $R_n$. $Q_n$ can be decomposed into Hadamard gates and controlled phase gates, with definitions of each at the bottom.}
    \label{fig:qft_circ}
\end{figure}
To present our main theorem, we start with the definition of the QFT. The QFT is usually defined as the Fourier transform over the finite cyclic group $\mathbb{Z}_{2^n}$ represented through amplitudes of a quantum state, or simply the discrete Fourier transform on a quantum state. Specifically, in the computational basis of qubits, it can be expressed as the unitary matrix
\begin{equation}
    F_n = \frac{1}{\sqrt{2^n}} \sum_{q=0}^{2^n-1} \sum_{q'=0}^{2^n-1} e^{i2\pi q q'/2^n}| q \rangle \langle q' |
    \label{eq:QFT_matrix_def}
\end{equation}
where $q$ the decimal representation of the bits $q_1, q_2, q_3,...$ i.e. $q = q_1 q_2...q_n$ and $q_i \in \{0, 1\}$. It is often more convenient to express $F_n$ as a linear map that transforms the qubit product state $|q_1 q_2...q_n\rangle$ into another product state,
\begin{multline}
    F_n|q_1 q_2 ... q_n\rangle = \frac{1}{\sqrt{2^n}} (|0\rangle + e^{2\pi i 0.q_n}|1\rangle) ~ \otimes \\
    (|0\rangle + e^{2\pi i 0.q_{n-1}q_n}|1\rangle) \otimes... (|0\rangle + e^{2\pi i 0.q_1 q_2 ...q_n}|1\rangle)
    \label{eq:QFT_def}
\end{multline}
where $0.q_1...q_n \equiv q_1/2 + ... + q_n/2^n$. This product representation of the linear map gives a natural way to construct the quantum circuit of the QFT by forming the decomposition $F_n = R_n Q_n$ \cite{nielsen_chuang_2010}, where $R_n$ is the bit reversal operator $R_n|q_1 ... q_n\rangle = |q_n ... q_1\rangle$ and $Q_n$ is the operator that transforms the product state in the following way:
\begin{multline}
    Q_n|q_1 q_2 ... q_n\rangle = \frac{1}{\sqrt{2^{n}}} (|0\rangle + e^{2\pi i 0.q_1 q_2...q_n}|1\rangle) ~ \otimes \\
    (|0\rangle + e^{2\pi i 0.q_{2}...q_n}|1\rangle) \otimes... (|0\rangle + e^{2\pi i 0.q_n}|1\rangle)
    \label{eq:rQFT_def}
\end{multline}
which arises naturally from the Fourier transform's recursive structure and whose gate components can be extracted easily from the product representation. See Fig.~\ref{fig:qft_circ} for an example of the QFT's circuit. In theory, $F_n$ and $Q_n$ are often considered equivalent since qubit ordering can be tracked classically. However, in practical implementations, different amounts of resources are required to implement the two operators, due to qubit architecture constraints (e.g. connectivity). For example, when restricted to nearest-neighbor connectivity, $F_n$ can be implemented in linear-depth \cite{Fowler_04, Maslov_07} while $Q_n$ may require more depth; if one allows all-to-all connectivity and ancilla qubits, both can be implemented in log-depth with very high accuracy \cite{Cleve_00}, but $Q_n$ contains much fewer gates. However, as said in the introduction, their entanglement structure defers a lot, with $Q_n$ having Schmidt coefficients decaying exponentially while $F_n$ has uniform Schmidt coefficients. This implies in situations where entanglement plays an important role, the two operators will behave very differently.

To understand the Schmidt coefficients of an operator and why they are important, it is necessary to introduce the operator Schmidt decomposition in the context of unitary operators:
\begin{definition}
     The \textup{operator Schmidt decomposition} of a unitary operator $U$ acting on a bipartite space $\mathcal{A} \otimes \mathcal{B}$ with dimension $dim(\mathcal{A\otimes \mathcal{B}}) = dim(\mathcal{A})dim(\mathcal{B}) = N$ is a decomposition of the form
    \begin{equation}
        U = \sqrt{N} \sum_{k=0}^{\chi-1} \sigma_{k} A_k \otimes B_k
        \label{eq:U_sd}
    \end{equation}
    where $\{A_k\}$ and $\{B_k\}$ are operators acting on $\mathcal{A}$ and $\mathcal{B}$ respectively, each satisfying $\Tr(A_k^\dagger A_{k'}) = \delta_{k,k'}$ and $\Tr(B_k^\dagger B_{k'}) = \delta_{k,k'}$, with $\delta_{k,k'}$ being the Kronecker delta. The constant $\chi$ is the Schmidt rank and is at most $min(dim(\mathcal{A})^2, dim(\mathcal{B})^2)$. The set of non-negative numbers $\{\sigma_k\}$ are called Schmidt coefficients, which in our case are organized in descending order and are $l^2$-normalized, i.e. $\sigma_0 \ge \sigma_1 \ge \dots$ and $\sum \sigma_k^2 = 1$.
\end{definition}
Equivalently, the operator Schmidt decomposition can be interpreted as the Schmidt decomposition of the vectorized operator $|U\rrangle$. The resulting state has norm $\sqrt{\Tr(U^\dagger U)} = \sqrt{N}$ and thus there is an extra $\sqrt{N}$ scalar in Eq.~\eqref{eq:U_sd}. The Schmidt coefficients represent the operator's expressibility as an MPO, which will be shown in more detail in Section~\ref{section:TN_sim_of_QFT}. Additionally, it is also a rough estimation of the operator's entanglement generation on a state, which will be discussed in Section~\ref{subsection:area_law} and Appendix~\ref{appendix:opEE_EP}.

The exact operator Schmidt decomposition of $F_n$ has been calculated in \cite{Nielsen_2003} and \cite{Tyson_2003}, where both papers showed that the Schmidt coefficients are uniform. This suggests the $F_n$ is maximally entangled and its MPO representation requires a bond dimension growing exponentially with the number of qubits. Through simulations, however, we noticed the core part of $F_n$, the operator $Q_n$, has Schmidt coefficients decaying exponentially quickly. This was previously and independently discovered in \cite{Woolfe_2014}, where the authors gave numerical evidence that the Schmidt coefficients decay very fast and the operator can be compressed to a constant-bond-dimension MPO with exponentially small error. However, no analytic analysis on the upper bounds of the Schmidt coefficients was given, and the underlying reasons for this low entanglement property were not studied. In this paper, we fill this gap by proving the following theorem.

\begin{theorem}
    \textup{(Exponential decay of the QFT's Schmidt coefficients).} Consider the quantum Fourier transform $Q_n$ defined in Eq.~\eqref{eq:rQFT_def} with the order of qubits at the output reversed. We partition the Hilbert space into $\mathcal{A}$ with qubits $1$ to $j$ and $\mathcal{B}$ with qubits $j+1$ to $n$, so that $Q_n$ has the operator Schmidt decomposition
    \begin{equation}
        Q_n = \sqrt{2^n} \sum_{k=0}^{\min(2^j,2^{n-j}) - 1} \sigma_{n,j}^{k} A_{n,j}^{k} \otimes B_{n,j}^{k}
        \label{eq:rQFT_sd}
    \end{equation}
    It then follows that as $n \to \infty$, for $k \ge 2$, $ \sigma_{n,j}^{k}$ has upper bound
    \begin{equation}
        \tilde{\sigma}_{j}^{k} = \lim_{n \to \infty} \sigma_{n,j}^{k} \, \leq \, \frac{1}{\sqrt{k}} \exp \left(-\frac{2k+1}{2}\log\left(\frac{4k+4}{e\pi}\right)\right)
        \label{eq:rQFT_sd_bound}
    \end{equation}
    Additionally, the following identity holds:
    \begin{equation}
        \sum_{k=0}^{k_0} \left(\sigma_{n+1,j}^{k}\right)^2 \leq \sum_{k=0}^{k_0} \left(\sigma_{n,j}^{k}\right)^2
        \label{eq:majorization}
    \end{equation}
    for any $n$, $j$ and $k_0$.
    \label{theorem:rQFT_low}
\end{theorem}
We will prove this theorem in Section~\ref{subsection:QFT_LE_proof}. Roughly in words, Eq.~\eqref{eq:rQFT_sd_bound} says as the qubit number goes to infinity, each Schmidt coefficient is still bounded by a quantity decaying (more than) exponentially fast. Note this bound is also independent of $j$, the partition of the system. Eq.~\eqref{eq:majorization} is usually referred to as \emph{majorization} for eigenvalues of two matrices, and it says the smaller the system size, the more concentrated at the beginning the QFT's Schmidt coefficients are. This makes sense intuitively because one would not expect a smaller QFT to contain more entanglement. A direct Corollary of Theorem~\ref{theorem:rQFT_low} is the following main statement of our paper.
\begin{corollary} 
    \textup{(Constant entanglement in the QFT).} For any number of qubits $n$ and partition $j$, the operator R\'enyi entropy of the QFT operator $Q_n$, defined as 
    \begin{equation}
        H_{\alpha}(n, j) = \frac{1}{1-\alpha} \log\left(\sum_{k}\left(\sigma_{n,j}^k\right)^{2\alpha}\right)
    \label{eq:op_renyi_entropy}
    \end{equation}
    is bounded by a constant for all $\alpha > 0$.
\end{corollary}
The proof follows directly from plugging the bounds from Eq.~\eqref{eq:rQFT_sd_bound}, as the sum converges for any fixed $\alpha > 0$, which implies $H_\alpha(\infty, j)$ is bounded by a constant. The majorization condition Eq.~\eqref{eq:majorization} ensures $H_\alpha(n, j) \leq H_\alpha(n+1, j) \leq H_\alpha(\infty, j)$, as the R\'enyi entropy is Schur concave.

\subsection{Proof of the Main Theorem}
\label{subsection:QFT_LE_proof}
\begin{figure}[tb]
    \centering
    \includegraphics[width=1\columnwidth]{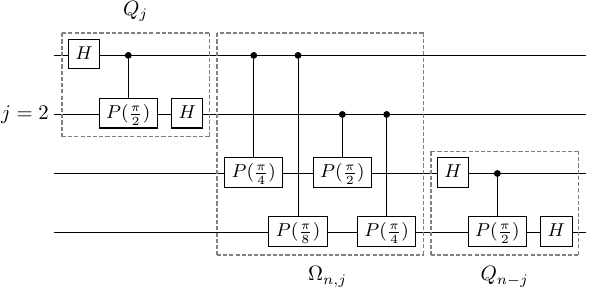}
    \caption{A 4-qubit example of the genearlized recurisve circuit of $Q_n$, i.e. $\Omega_{n,j}$ sandwiched by two smaller QFT circuits, $Q_j$ and $Q_{n-j}$. While in this example we choose $j=n/2=2$, this decomposition works for any $j$.}
    \label{fig:qft_circ_rearrange}
\end{figure}
In this subsection, we prove Theorem~\ref{theorem:rQFT_low}. We first show that the normalized Schmidt coefficients of $Q_n$ are equivalent to singular values of submatrices of $F_n$. Then we show this relates to a well-known problem in signal processing called the \emph{spectral concentration problem}. The upper bounds of solutions to this problem have been studied extensively, which we will use to prove Theorem~\ref{theorem:rQFT_low}.

Consider an $n$-qubit quantum circuit of the QFT operator $Q_n$ shown in Fig.~\ref{fig:qft_circ}. If we partition the circuit into the top part with qubits $1$ to $j$, and the bottom part with qubits $j+1$ to $n$, one can always push gates in the top part as left as possible (acting first on the state) to form a $j$-qubit QFT, and push gates in the bottom as right as possible (acting last) to form a $(n-j)$-qubit QFT, with some remaining controlled phase gates in the center. A circuit diagram illustration is shown in Fig.~\ref{fig:qft_circ_rearrange}. This results in the decomposition
\begin{equation}
    Q_n = \left(I_j \otimes Q_{n-j}\right) \Omega_{n,j} \left(Q_{j} \otimes I_{n-j}\right)
    \label{eq:QFT_decomp}
\end{equation}
where $I_{j}$ is the $2^j \times 2^j$ identity matrix and $\Omega_{n,j}$ is the diagonal unitary matrix defined as
\begin{equation}
    \Omega_{n,j}=\prod_{l=1}^j \prod_{m=j+1}^{n} P_{l,m}\Big(\frac{\pi}{2^{m-l}}\Big)
\end{equation} 
where $P_{l,m}(\theta)$ is the controlled phase gate with qubit $l$ as control and qubit $m$ as target. All phase gates commute with each other so the order of products here is not important. A $P_{l,m}(\theta)$ is part of $\Omega_{n,j}$ if and only if it goes across the partition between qubit $j$ and $j+1$. Equivalently, one can interpret this decomposition as a recursive construction scheme for $Q_n$ with $Q_1=H$, and was termed the \emph{generalized QFT circuit} \cite{Cleve_00}. For a detailed proof of the decomposition, see Appendix~\ref{appendix:qft_decomposition}. Because the two smaller QFT circuits $Q_{j}$ and $Q_{n-j}$ are unitary matrices acting only on one side of the partition, they can be dropped without affecting the Schmidt-coefficients of the QFT partitioned at $j$. Thus:
\begin{equation}
    \text{Sch}_j(Q_n) = \text{Sch}_j(\Omega_{n,j})
\end{equation}
where $\text{Sch}_j(A)$ denotes the set of Schmidt coefficients calculated by partitioning the operator $A$ between qubits $j$ and $j+1$. To calculate and simplify the matrix elements of $\Omega_{n,j}$, we observe that in the subspace spanned by qubits $l$ and $m$, $P_{l,m}(\theta)$ can be represented as the linear map
\begin{equation}
    P_{l,m}(\theta)|q_lq_m\rangle = e^{i \theta q_lq_m} |q_lq_m\rangle
\end{equation} 
Therefore, $\Omega_ {n,j}$ can be represented as the matrix
\begin{multline}
    \Omega_{n,j} = \sum_{q} \exp (i\sum_{l=1}^{j} \sum_{m=j+1}^{n} q_l q_m \frac{\pi}{2^{m-l}}) \times \\ |q_1...q_n\rangle \langle q_1...q_n|
\end{multline}
where $\sum_q$ sums over all bit-strings $q = q_1 q_2...q_n$. Next we partition the bit-string $q$ and define $x=x_1...x_j=q_j...q_1$ and $x'=x'_1...x'_{n-j}=q_{j+1}...q_{n}$, which allows $\Omega_{n,j}$ to be rewritten as
\begin{multline}
    \Omega_{n,j} = \sum_{x,x'} \exp (i\sum_{l=1}^{j} \sum_{m=1}^{n-j} x_{l} x'_{m} \frac{\pi}{2^{l+m-1}}) \times \\ |x_j...x_1\rangle \langle x_j...x_1| \otimes |x_1'...x_{n-j}'\rangle \langle x_1'...x_{n-j}'|
    \label{eq:Omega_as_xxp}
\end{multline}
Then if we explicitly write out the binary expansion of $x$ and $x'$ as sums:
\begin{equation}
\begin{dcases}
    x = \sum_{l=1}^{j}{x_{l}2^{j-l}} = 2^j\sum_{l=1}^{j}{\frac{x_{l}}{2^{l}}}\\
    x' = \sum_{m=1}^{n-j}{x'_{m}2^{n-j-m}} = 2^{n-j}\sum_{m=1}^{n-j}{\frac{x'_{m}}{2^{m}}}
\end{dcases}  
\end{equation}
and substitute back to Eq.~\eqref{eq:Omega_as_xxp}, $\Omega_{n,j}$ can be further simplified to
\begin{equation}
    \Omega_{n,j} = \sum_{x,x'} e^{i2\pi x x'/2^n} |x^r\rangle \langle x^r| \otimes |x'\rangle \langle x'|
\end{equation}
where $x^r$ is the bit-reversed representation of $x$, i.e. $|x^r\rangle = R_j|x\rangle = |x_j...x_1\rangle$. The Schmidt coefficients of  $\Omega_{n,j}$ can be then obtained by doing the reshaping $|x^r\rangle \langle x^r| \otimes |x'\rangle \langle x'| \rightarrow |x^r x^r\rangle \langle x'x'|$ and calculating the singular values:
\begin{equation}
\begin{split}
    \text{Sch}_j (\Omega_{n,j}) &= \text{Sch}_j \left(\sum_{x,x'} e^{i2\pi x x'/2^n} |x^r\rangle \langle x^r| \otimes |x'\rangle \langle x'| \right)\\
    &= \sigma \left( \frac{1}{\sqrt{2^n}} \sum_{x,x'} e^{i2\pi x x'/2^n} |x^rx^r\rangle \langle x'x'|\right)\\
    &= \sigma \left( \frac{1}{\sqrt{2^n}}  \sum_{x,x'} e^{i2\pi x x'/2^n} |x\rangle \langle x'|\right)
\end{split}
\label{eq:reshape_Omega}
\end{equation}
where $\sigma(A)$ denotes the non-zero singular values of matrix $A$. The extra constant $1/\sqrt{2^n}$ comes from the normalization of Schmidt coefficients. The third line in Eq.~\eqref{eq:reshape_Omega} removes duplicated dimensions $|x^r x^r\rangle\langle x' x'| \rightarrow |x^r\rangle\langle x'|$, which removes zero-valued rows and columns, and then reverses the bitstring back $|x^r\rangle\langle x'| \rightarrow |x\rangle\langle x'|$, which are permutations of rows. Both actions preserve non-zero singular values. The resulting matrix is in fact the $2^j \times 2^{n-j}$ top-left submatrix of $F_n$, which we denote as $F_{n,j}$:
\begin{equation}
\begin{split}
    F_{n,j} &= F_{n}[0:2^j-1,0:2^{n-j}-1]\\
    &= \frac{1}{\sqrt{2^n}} \sum_{x=0}^{2^j-1} \sum_{x'=0}^{2^{n-j}-1}  e^{i2\pi x x'/2^n} |x\rangle \langle x'|
\end{split}
\end{equation} 
and then it follows that $\text{Sch}_j(\Omega_{n,j}) = \sigma(F_{n,j})$. Because $F_{n}$ is a symmetric matrix, a property $F_{n,j}$ has is $\left(F_{n,j}\right)^T = F_{n,n-j}$, thus the Schmidt coefficients on cut $j$ and $n-j$ are exactly the same.

It turns out the submatrix of the QFT matrix $F_n$ was previously known to be approximately low rank, in particular, it was studied through the \emph{spectral concentration problem}, which are introduced in detail in Appendix~\ref{appendix:spectral_concentration}. To see the connection, consider the positive-definite matrix $T_{n,j} = F_{n,j}\left(F_{n,j}\right)^\dagger$, which has matrix elements
\begin{equation}
\begin{split}
    T_{n,j} = & \frac{1}{2^n} \sum_{x,y,x',x''} e^{i2\pi (xx'-x''y)/2^n} |x\rangle \langle x'|x''\rangle \langle y| \\
    = & \frac{1}{2^n} \sum_{x,y} \sum_{x'=0}^{2^{n-j}-1} e^{i2\pi (x-y) x'/2^n} |x\rangle \langle y|
\end{split}
\label{eq:A_elements}
\end{equation}
We can apply the summation equation of the geometric series to remove the sum over $x'$ and simplify the matrix elements
\begin{equation}
\begin{split}
    T_{n,j} = & \frac{1}{2^n} \sum_{x,y} \frac{1-e^{i2\pi(x-y)2^{n-j}/2^{n}}}{1-e^{i2\pi(x-y)/2^{n}}} |x\rangle \langle y|\\
    = & \frac{1}{2^n} \sum_{x,y} \omega_{n,j}^{x-y}\frac{\text{sin}(\pi(x-y)/2^j)}{\text{sin}(\pi(x-y)/2^n)} |x\rangle \langle y|\\
\end{split}
\label{eq:A_elements_simplified}
\end{equation}
where $\omega_{n,j}^{x-y} = e^{i\pi(x-y)(2^{-j}-2^{-n})}$. The $(k+1)$-th eigenvalue $\lambda_{n,j}^{k}$  of $T_{n,j}$ is thus obtained by solving the equation
\begin{equation}
    \begin{split}
        T_{n,j} |u_{n,j}^{k}\rangle &= \lambda_{n,j}^{k} |u_{n,j}^{k}\rangle\\
        \frac{1}{2^n}\sum_{y=0}^{2^j-1} \omega_{n,j}^{x-y}\frac{\text{sin}(\pi(x-y)/2^j)}{\text{sin}(\pi(x-y)/2^n)} u_{n,j}^{k,y} &= \lambda_{n,j}^{k} u_{n,j}^{k,x}\\
        \frac{1}{2^n}\sum_{y=0}^{2^j-1} \frac{\text{sin}(\pi(x-y)/2^j)}{\text{sin}(\pi(x-y)/2^n)} \omega_{n,j}^{-y} u_{n,j}^{k,y} &= \lambda_{n,j}^{k} \omega_{n,j}^{-x} u_{n,j}^{k,x}\\
        \frac{1}{2^n}\sum_{y=0}^{2^j-1} \frac{\text{sin}(\pi(x-y)/2^j)}{\text{sin}(\pi(x-y)/2^n)} v_{n,j}^{k,y} &= \lambda_{n,j}^{k} v_{n,j}^{k,x}
    \end{split}
    \label{eq:A_eig}
\end{equation}
where $u_{n,j}^{k,x}$ denotes the $(x+1)$-th element of the vector $|u_{n,j}^{k}\rangle$ and $v_{n,j}^{k,x} = \omega_{n,j}^{-x} u_{n,j}^{k,x}$. The last equation in Eq.~\eqref{eq:A_eig} belongs to a family of eigenvalue equations studied extensively in Fourier analysis, where the eigenvectors are known as the \emph{periodic discrete prolate spheroidal sequences} (P-DPSSs) \cite{Xu_1984}. The general P-DPSS eigenvalue equations are
\begin{equation}
    \frac{1}{N}\sum_{y=0}^{N'-1} \frac{\text{sin}\left(2W\pi(x-y)\right)}{\sin (\pi(x-y)/N)} v_{N,N',W}^{k,y} = \lambda_{N, N',W}^{k} v_{N,N',W}^{k,x}
    \label{eq:PDPSS}
\end{equation}
where in our case $N=2^n$, $N'=2^j$ and $2W=1/2^j$. The P-DPSS eigenvalues have a special clustering behavior: slightly fewer than $2N'W$ eigenvalues are very close to or equal to 1, slightly fewer than $N'-2N'W$ eigenvalues are very close to or equal to 0, and very few eigenvalues are not near 1 or 0 \cite{Karnik_2021}. For Eq.~(\ref{eq:A_eig}), $2N'W=2^j/2^j=1$, suggesting nearly all eigenvalues besides the leading ones are close to or equal to 0. Eigenvalues equal to 0 arise from the symmetry $\left(F_{n,j}\right)^T = F_{n,n-j}$ we discussed: $T_{n,j} = F_{n,j}\left(F_{n,j}\right)^\dagger$ has dimension $2^j \times 2^j$ but only rank $\min(2^j, 2^{n-j})$, thus when $j > n/2$ we have a non-empty null-space.

While P-DPSSs are widely used in signal processing, there are only a few results on the upper bounds of their eigenvalues \cite{Edelman_1998, Matthysen_2016, Zhu_2018}. In the literature, the bounds are either empirical or loose when $2N'W$ is small. However, a set of closely related eigenvalue equations is studied much more frequently, which is when $n \to \infty$. As $n \to \infty$, the last line of Eq.~\eqref{eq:A_eig} reduces to
\begin{equation}
    \sum_{y=0}^{2^j-1} \frac{\text{sin}(\pi(x-y)/2^j)}{\pi(x-y)} \tilde{v}_{j}^{k,y} = \tilde{\lambda}_{j}^{k} \tilde{v}_{j}^{k,x}
    \label{eq:DPSS_qft}
\end{equation}
whose eigenvectors $|\tilde{v}_{j}^{k}\rangle$ belong to a kind of vectors known as the \emph{discrete prolate spheroidal sequences} (DPSSs) \cite{Slepian_1978}. The general DPSSs $\tilde{v}_{N',W}^{k,y}$ come from solving the eigenvalue equation
\begin{equation}
    \sum_{y=0}^{N'-1} \frac{\text{sin}\left(2W\pi(x-y)\right)}{\pi(x-y)} \tilde{v}_{N',W}^{k,y} = \tilde{\lambda}_{N',W}^{k} \tilde{v}_{N',W}^{k,x}
    \label{eq:DPSS}
\end{equation} 
where again, in the context of the QFT, we set $N'=2^j$ and $2W=1/2^j$. DPSSs are solutions to the spectral concentration problems, and P-DPSSs can be interpreted as solutions to the discrete version of such problems. While DPSSs and their eigenvalues have not been solved analytically either, there are numerous results on the bounds of their eigenvalues \cite{Slepian_1978, Karnik_2019_1, Matthysen_2016, Karnik_2019_2, Zhu_2015, Boulsane_2020, Karnik_2021, Bonami_2021}. We specifically adopt techniques from \cite{Boulsane_2020, Bonami_2021} and prove in Appendix~\ref{appendix:DPSS_upperbounds} that for $k \ge 2$:
\begin{equation}
    \tilde{\sigma}_{j}^{k} = \sqrt{\tilde{\lambda}_{j}^{k}} \leq \frac{1}{\sqrt{k}} \exp \left(-\frac{2k+1}{2}\log\left(\frac{4k+4}{e\pi}\right)\right)
    \label{eq:rQFT_sd_bound_2}
\end{equation}
Therefore Eq.~\eqref{eq:rQFT_sd_bound} in Theorem~\ref{theorem:rQFT_low} is proved.

While this bound is only for the limiting case $n \to \infty$, in Appendix~\ref{appendix:spectral_dynamics} we show that, with a mild assumption, the bounds for any non-leading eigenvalues of Eq.~\eqref{eq:DPSS} can be directly applied to Eq.~\eqref{eq:PDPSS}, i.e.
\begin{equation}
    \begin{cases}
        \sigma_{n,j}^{0} \ge \tilde{\sigma}_{j}^{0} &\\
        \sigma_{n,j}^{k} \le \tilde{\sigma}_{j}^{k}, &k \ge 1 \\
    \end{cases}
\end{equation}
Thus the upper bounds for DPSSs in Eq.~\eqref{eq:rQFT_sd_bound_2} are also upper bounds for Schmidt coefficients of QFT. 

We now prove Eq.~\eqref{eq:majorization} in Theorem~\ref{theorem:rQFT_low}, i.e. the eigenvalues of $T_{n+1, j}$ ($\lambda_{n+1, j}^k$) and those of $T_{n, j}$ ($\lambda_{n, j}^k$) satisfy
\begin{equation}
    \sum_{k = 0}^{k_0} \lambda_{n+1,j}^k \leq \sum_{k = 0}^{k_0} \lambda_{n,j}^k
    \label{eq:T_majorization}
\end{equation}
for arbitrary $n$, $j$ and $k_0$.
We first expand $T_{n+1, j}$ as
\begin{equation}
\begin{split}
    T_{n + 1,j}[x, y] &= \frac{1}{2^{n+1}} \frac{\text{sin}(\pi(x-y)/2^j)}{\text{sin}(\pi(x-y)/2^{n+1})} \\
    &= \frac{1}{2} \frac{\text{sin}(\pi(x-y)/2^n)}{\text{sin}(\pi(x-y)/2^{n+1})} T_{n,j}[x, y] \\
    &= \cos (\pi (x-y)/2^{n+1}) T_{n,j}[x, y]
\end{split}
\end{equation}
which is a Hadamard product (denoted as $\circ$) between two matrices
\begin{equation}
    T_{n + 1,j} = C_{n + 1} \circ T_{n,j}
\end{equation}
where we define $C_n$ to have matrix elements:
\begin{equation}
    C_{n}[x, y] = \cos (\pi (x-y)/2^n) 
\end{equation}
One can verify that $C_n$ is a correlation matrix, i.e. it is positive-semidefinite (PSD) and its diagonals are all 1. This directly establishes the majorization relation
\begin{equation}
    T_{n + 1,j} \prec T_{n,j}
\end{equation}
because the Hadamard product with a correlation matrix is a doubly-stochastic linear map, which directly implies majorization \cite{Zhan_2002}. Therefore Theorem~\ref{theorem:rQFT_low} is proved.

\subsection{Relation to the Area Law}
\label{subsection:area_law}
In this subsection, we develop a more intuitive explanation for QFT's low entanglement. Specifically, we show that QFT has the same entanglement structure as the time evolution of a Hamiltonian with exponential decay of interaction, which is pseudo-local. Dynamics under such interactions are known to obey a variant of the area law \cite{Van_Acoleyen_2013, Gong_2017}, which bounds by a constant how much the entanglement entropy of a system can be changed. 

We first claim that removing $H$ gates from $Q_n$ does not change its entanglement structure. To see how, recall the equation introduced in Section \ref{subsection:QFT_LE_proof},
\begin{equation}
    \text{Sch}_j(Q_n) = \text{Sch}_j(\Omega_{n,j})
\end{equation}
one observes that since $\Omega_{n,j}$ only contains controlled phase gates, the $H$ gates do not play any role in $Q_n$'s Schmidt coefficients. More generally, if we denote the operator obtained from removing $H$ gates in $Q_n$ as $Q_n^{P}$, we can show that any reasonable bipartite entanglement measure is identical between $Q_n$ and $Q_n^{P}$; see Appendix~\ref{appendix:opEE_EP} for details. Therefore, the entanglement study of $Q_n^{P}$ can be directly applied to $Q_n^{P}$. Note this is entirely due to the QFT circuit's special structure, and generally single-qubit gates sandwiched by multi-qubit gates will change the entanglement structure of the operator.

Interestingly, $Q_n^{P}$ can be physically interpreted as the time-evolution operator of an exponentially decaying Hamiltonian over a time interval of $t=\pi$. To see how, we first express $Q_n^P$ as a product of phase gates:
\begin{equation}
    Q_n^{P} = \prod_{l=1}^{n-1}\prod_{m=l+1}^n P_{l,m}(\frac{\pi}{2^{m-l}})
\end{equation}
One can expand $P_{l,m}(\theta)$ in the subspace spanned by qubit $l$ and $m$ as
\begin{equation}
    P_{l,m}(\theta) = \exp(i\theta |1 \rangle\langle 1|_{l} \otimes |1 \rangle\langle 1|_{m})
\end{equation}
where $|1\rangle\langle1|_l$ is the projection to state $|1\rangle$ on qubit $l$. $Q_n^{P}$ can then be expressed as
\begin{equation}
    Q_n^{P} = \exp(i\pi\sum_{l=1}^{n-1}\sum_{m=l+1}^n \left(\frac{1}{2}\right)^{m-l} |1 \rangle\langle 1|_{l} \otimes |1 \rangle\langle 1|_{m})
\end{equation}
which arises from time evolving by $\pi$ time units under the Hamiltonian
\begin{equation}
    H_n^P = -\sum_{l=1}^{n-1}\sum_{m=l+1}^n \left(\frac{1}{2}\right)^{m-l} |1 \rangle\langle 1|_{l} \otimes |1 \rangle\langle 1|_{m}
\end{equation}
Note that the projection operator $|1 \rangle\langle 1|$ is equivalent to the Pauli-Z operator shifted and rescaled, so the Hamiltonian $H_n^P$ can also be physically interpreted as $Z$-$Z$ interaction on each pair of sites with exponentially decaying correlations. Such interaction is considered as short-range, and the sum of interactions across any partition converges as the system size increases, thus so does the entanglement entropy. Specifically, in \cite{Gong_2017} the authors prove that for a quantum state on a $D$-dimensional finite or infinite lattice, if it evolves in time under a Hamiltonian with interactions decaying faster than $1/r^{D+1}$ where $r$ is the distance between two sites, the entanglement entropy of the state with respect to a subregion cannot change faster than a rate proportional to the size of the subregion's boundary. For the case of $Q_n^{P}$ on a 1D qubit chain, the interaction decays as $1/2^r$ which is exponentially faster than $1/r^2$. Therefore, the entanglement rate is bounded by a very small constant, and the amount of entanglement generated through $Q_n^{P}$ on any state $|\psi\rangle$ does not grow with the number of qubits, i.e.
\begin{equation}
    \Delta S_j = \left| S_j(Q_n^{P}|\psi\rangle) - S_j(|\psi\rangle) \right| \leq O(1)
    \label{eq:delta_S}
\end{equation}
where $S_j$ is the von Neumann entropy calculated at partition $j$. See Appendix~\ref{appendix:opEE_EP} for details. 

While this result does not directly translate to Schmidt coefficients of $Q_n^{P}$ (or those of $Q_n$), we note the operator Schmidt decomposition can be reinterpreted as the Schmidt decomposition of the state $Q_n^{P}(|\alpha\rangle\otimes|\beta\rangle)$, where $|\alpha\rangle$ is a maximally entangled state on system $\mathcal{A}$ and an ancilla $\mathcal{R_A}$ with $\text{dim}(\mathcal{R_A}) = \text{dim}(\mathcal{A})$, and $|\beta\rangle$ is a maximally entangled state on system $\mathcal{B}$ and an ancilla $\mathcal{R_B}$ with $\text{dim}(\mathcal{R_B}) = \text{dim}(\mathcal{B})$. Therefore, the Schmidt coefficients of $Q_n^{P}$ arise from setting $|\psi\rangle = |\alpha\rangle\otimes|\beta\rangle$ in Eq.~\eqref{eq:delta_S}, which indicates that they must decay fast enough so they converge under the function $S_j$ and thus $\Delta S$ is a constant. Indeed, our Theorem~\ref{theorem:rQFT_low} is a stronger statement, which shows that the Schmidt coefficients decay exponentially fast, so $\Delta S$ is not only bounded by a constant but also extremely small.

\section{Matrix Product Operator Form of the QFT}
\label{section:TN_sim_of_QFT}
The low-entanglement property of the QFT naturally raises the question of whether it is efficient to simulate it classically. To answer it, we consider the tensor network simulations of the QFT in this section. We briefly review the definition of tensor networks, tensor diagrams, and the matrix product operator (MPO). Then, we introduce a tensor network representation of the QFT's circuit, and show how to compress it into a bond-dimension-$\chi$ MPO with $O(n e^{-\chi \log(\chi/3)}/\sqrt{\chi})$ error in $O(\chi^3 n^2)$ time. In the next section, we will apply this MPO in the classical setting and compare it against the fast Fourier transform (FFT).

\subsection{Tensor Networks and the Matrix Product Operators}
\label{subsection:TN_MPO}
We briefly review the definition of tensor networks, tensor diagrams, and the MPO. In our context, a \emph{tensor} is a $d$-dimensional array of complex numbers, with $d$ being the \emph{order} of the tensor or its number of indices. The $contraction$ of a set of tensors is the sum of all the possible values of indices shared by tensors; for example,
\begin{equation}
    A_{\alpha}^{\gamma} = \sum_{\beta} B_{\beta}^{\alpha} C_{\beta}^{\gamma}
    \label{eq:TN_contraction}
\end{equation}
represent contracting tensors $B$ and $C$, where $\alpha, \beta, \gamma$ are index variables and $\alpha, \gamma$ are \emph{open indices}. Indices are put as either superscripts or subscripts, and in our context, both serve the same meaning. A \emph{tensor network} is then a network of tensors being contracted in certain ways, with some (or none) of the indices left open. A \emph{tensor diagram} is the diagrammatic representation of a tensor network following two basic rules: firstly, tensors are notated by shapes (e.g. squares, circles, triangles...), and tensor indices are notated by lines emanating from these shapes; secondly, connecting two index lines implies a contraction. For example, Eq.~\eqref{eq:TN_contraction} can be represented by the diagram
\begin{equation}
    \input{TN_example.tikz}
    \label{eq:TN_example}
\end{equation}
The diagram represents the same tensor network if shapes are rearranged or rotated and lines are bent, as long as contracted indices remain the same. Additionally, in our context, crossing lines do not intersect, i.e. one line can be thought of as out of the plane. To demonstrate the above points, the following two tensor diagrams are considered equivalent:
\begin{equation}
    \input{TN_example_bend.tikz}
    \label{eq:TN_example_bend}
\end{equation}
In physics, a commonly used tensor network structure is the matrix product state (MPS), which in our context is a contracted chain of order-3 tensors with order-2 tensors on the boundary. We can generalize the matrix product representation into operator space, which we call the matrix product operator (MPO). It is analogous to MPS except that each tensor has one more input index to represent an operator instead of a state. More specifically, they are defined as follows:
\begin{definition}
    A \textup{matrix product operator (MPO)} representation of an order-$2n$ tensor $O$ is a tensor network of the form
    \begin{equation}
        O_{x_1,...x_n}^{y_1,...y_n} = \sum_{b_1,...b_{n-1}} T[1]_{b_1}^{x_1, y_1} T[2]_{b_1, b_2}^{x_2, y_2} ... T[n]_{b_{n-1}}^{x_{n}, y_{n}}
    \end{equation}
    where $\{T[k]\}$ are referred as the tensors on the $k$-th site, $\{b_k\}$ are referred as \textup{bond indices}, and $\{x_k\}, \{y_k\}$ are referred as \textup{site indices}. The dimension of bond indices is usually denoted as \textup{bond dimension}. Diagrammatically, an MPO can be represented as
    \begin{equation}
        \input{MPO.tikz}
    \end{equation}
\end{definition}
In certain cases, the tensors on the MPO may satisfy the \emph{isometric} condition, which is a useful property that allows us to construct the \emph{canonical form} of the MPO. See the following two definitions for details.
\begin{definition}
    A tensor $T$ is \textup{isometric} with respect to input space formed by a subset of its indices $\{x_1, ..., x_m\}$ and output space formed by the rest of indices if and only if the following identity holds:
    \begin{equation}
         \input{isometric_tensor.tikz}
    \end{equation}
    where $T^*$ is the conjugate of $T$ (ie. the same tensor except that all the elements are complex conjugated). The indices $x_1', x_2',...$ of $T^*$ correspond to $x_1, x_2, ...$ of $T$ respectively, where the rest of the indices are contracted and are matched between $T$ and $T^*$.
\end{definition}

\begin{definition}
    An MPO is in \textup{canonical form} if it takes the form
    \begin{equation}
        \begin{split}
            \sum_{b_1,...,b_{n-1}} U[1]^{x_1, y_1}_{b_1}U[2]^{x_2, y_2}_{b_1, b_2}...T[c]^{x_{c},y_{c}}_{b_{c-1}, b_{c}}...\times\\
            V[n-1]^{x_{n-1}, y_{n-1}}_{b_{n-2}, b_{n-1}}V[n]^{x_{n}, y_{n}}_{b_{n-1}}
        \end{split}
    \end{equation}
    where tensors $
    \{U[k]\}$ satisfy the left (up) isometric condition
    \begin{equation}
        \sum_{x_{k}, y_{k}, b_{k-1}} {U^*}[k]_{b_{k-1}, b_k'}^{x_{k}, y_{k}} {U}[k]_{b_{k-1}, b_k}^{x_{k}, y_{k}} = \delta_{b_k', b_{k}}
    \end{equation}
    and tensors $\{V[k]\}$ satisfy the right (down) isometric condition
    \begin{equation}
        \sum_{x_{k}, y_{k}, b_{k}} {V}[k]_{b_{k-1}, b_{k}}^{x_{k}, y_{k}} {V^{*}}[k]_{b_{k-1}', b_{k}}^{x_{k}, y_{k}} = \delta_{b_{k-1}, b_{k-1}'}
    \end{equation}
    and $T[c]$ is a general tensor and is called the \textup{orthogonality center}. If $c=1$, we call this MPO \textup{left (up)-canonical}, and if $c=n$ we call it \textup{right (down)-canonical}. Diagrammatically, an MPO in canonical form can be represented as
    \begin{equation}
        \input{MPO_ortho.tikz}
    \end{equation}
    with $\{U[k]\}, \{V[k]\}$ satisfying
    \begin{equation}
        \input{ortho.tikz}
    \end{equation}
\end{definition}
The canonical form of MPSs and MPOs are useful, mostly because by factorizing the orthogonality center one can extract the Schmidt coefficients of the full system with respect to partition next to the orthogonality center. In fact, together with the QFT tensor network introduced in the next subsection, we show in Appendix~\ref{appendix:schmidt_from_QFT_TN} an alternative approach to obtain the QFT's Schmidt coefficients using the canonical form. MPSs in the canonical form are key ingredients for the density matrix renormalization group (DMRG) algorithm to work \cite{White_1992, White_1993, Schollwock_2005}. Additionally, the canonical forms are useful for MPS/MPO contractions. For example, the fitting algorithm employs the canonical form to contract in a DMRG-like style \cite{Verstraete_2004}, and the zip-up algorithm heuristically uses the orthogonality center to stabilize the contraction \cite{Stoudenmire_2010}.

\subsection{The QFT Tensor Network and the QFT-MPO}
\label{subsection:QFT_TN}
In this subsection, we introduce a tensor network (TN) representation of the QFT derived from the quantum circuit and how to turn it into a bond-dimension-$\chi$ MPO in $O(\chi^3 n^2)$ time with error $O(n e^{-\chi \log(\chi/3)}/\sqrt{\chi})$. 

We start by reviewing how quantum circuit diagrams can be exactly reinterpreted as tensor network diagrams. Consider the definition of a controlled phase gate, 
\begin{equation}
    \input{CP_definition.tikz}
    \label{eq:CP_definition}
\end{equation}
One can rewrite it as an inner product of two operator-valued vectors
\begin{equation}
    \begin{pmatrix}
        |0\rangle\langle 0| & |1\rangle\langle 1|
    \end{pmatrix}
    \begin{pmatrix}
    I\\P(\theta)
    \end{pmatrix}
\end{equation}
This decomposition can be diagrammatically represented as breaking the controlled phase gate into two order-3 tensors, and thus forming a two-site MPO
\begin{equation}
    \input{CP_to_tensor.tikz}
    \label{eq:CP_to_tensor}
\end{equation}
The tensor on the top is an order-3 \emph{copy tensor}
\begin{equation}
   \input{copy.tikz}
   \label{eq:copy}
\end{equation}
where $\delta_{a,b,c}=\delta_{a,b} \delta_{b,c}$ and $\delta_{a,b}$ is the Kronecker delta. The tensor on the bottom is an order-3 \emph{phase tensor}
\begin{equation}
    \input{3-phase.tikz}
    \label{eq:3-phase}
\end{equation}
In the two above equations, the horizontal indices $x_1$ and $x_1'$ run over the indices of operators $|0\rangle\langle 0|$, $|1\rangle\langle 1|$, $I$ and $P(\theta)$, and the vertical index $x_1''$ and $x_2$ run over the operator-valued vectors. More generally, vector or matrix indices (forming the tensor network) are the vertical indices in the tensor diagram and the horizontal indices are for operators. Note that here we are reinterpreting the vertical lines: in a quantum circuit diagram, the vertical lines are used merely for indicating the control-target relations, but here we consider them as tensor indices, just like the horizontal tensor indices for operators acting on qubits. Such reinterpretation is often used when considering quantum circuits as tensor networks \cite{Biamonte_2017}. 

We then show how a series of controlled phase gates can be turned into a single MPO with bond dimension two. To start with, we introduce an identity of copy tensors:
\begin{equation}
    \input{copy_simple_identity.tikz}
\end{equation}
In general, an order-$n$ copy tensor can be decomposed into an arbitrary number of copy tensors contracted together in arbitrary order, as long as the total number of open indices remains the same. For details see Appendix~\ref{appendix:schmidt_from_QFT_TN}. This allows us to deform the tensor diagram of two adjacent controlled phase gates as follows,
\begin{equation}
    \input{CP_series_to_tensor.tikz}
\end{equation}
If we contract the copy tensor and order-3 phase tensor in the center into one tensor:
\begin{equation}
    \input{4-phase_derivation.tikz}
\end{equation}
for which we define as the order-4 phase tensor
\begin{equation}
    \input{4-phase.tikz}
    \label{eq:4-phase}
\end{equation}
The series of controlled phase gates in the QFT circuit can be then turned into the MPO
\begin{equation}
    \input{cadder.tikz}
    \label{eq:cadder}
\end{equation}
corresponding to the decomposition
\begin{equation}
    \begin{pmatrix}
        |0\rangle\langle 0| & |1\rangle\langle 1|
    \end{pmatrix}
    \begin{pmatrix}
        I & 0 \\ 0 & P(\frac{\pi}{2})
    \end{pmatrix}
    \begin{pmatrix}
        I & 0 \\ 0 & P(\frac{\pi}{4})
    \end{pmatrix}
    ...
    \begin{pmatrix}
    I\\P(\frac{\pi}{2^{n-1}})
    \end{pmatrix}
\end{equation}
which we will refer as the \emph{phase MPOs}. 

It is easy to verify that the copy tensor on the first (i.e. top)  site is an isometric tensor
\begin{equation}
    \input{3-copy_isometric.tikz}
    \label{eq:3-copy_isometric}
\end{equation}
and the order-4 phase tensor is isometric up to a scalar 2 in both upward and downward directions:
\begin{equation}
    \input{4-phase_isometric.tikz}
    \label{eq:4-phase_isometric}
\end{equation}
However, the order-3 phase tensor at the last site is not isometric, i.e.
\begin{equation}
    \input{3-phase_not_isometric.tikz}
    \label{eq:3-phase_not_isometric}
\end{equation}
Thus up to a scaling factor of $2^{n-2}$, we can view the phase MPO as having the orthogonality center on the last site.

We also introduce the $H$ tensor
\begin{equation}
    \input{h-tensor.tikz}
    \label{eq:h-tensor}
\end{equation}
which is identical to an $H$ gate. The QFT circuit of $Q_n$ can thus be expressed as a series of MPOs in Eq.~\eqref{eq:cadder} with $H$ tensors sandwiched between each, which we call the QFT tensor network (QFT-TN). As an example, the 4-qubit circuit $Q_4$ has the tensor diagram
\begin{equation}
    \input{QFT_example_4q.tikz}
    \label{eq:qft_example_4q}
\end{equation}
In some literature, this is exactly how the circuit is drawn, where the MPO in our tensor network represents controlled multi-qubit phase gates. The QFT-TN has many additional properties, for example in Appendix~\ref{appendix:schmidt_from_QFT_TN} we show one could obtain the Schmidt coefficients by simplifying the QFT-TN, and in Appendix~\ref{appendix:emergent_QFT_circuit} we show one could derive the QFT circuit from a more generalized QFT-TN that represents any submatrix of the QFT.

\begin{figure*}[tb]
    \centering
    \includegraphics[width=2\columnwidth]{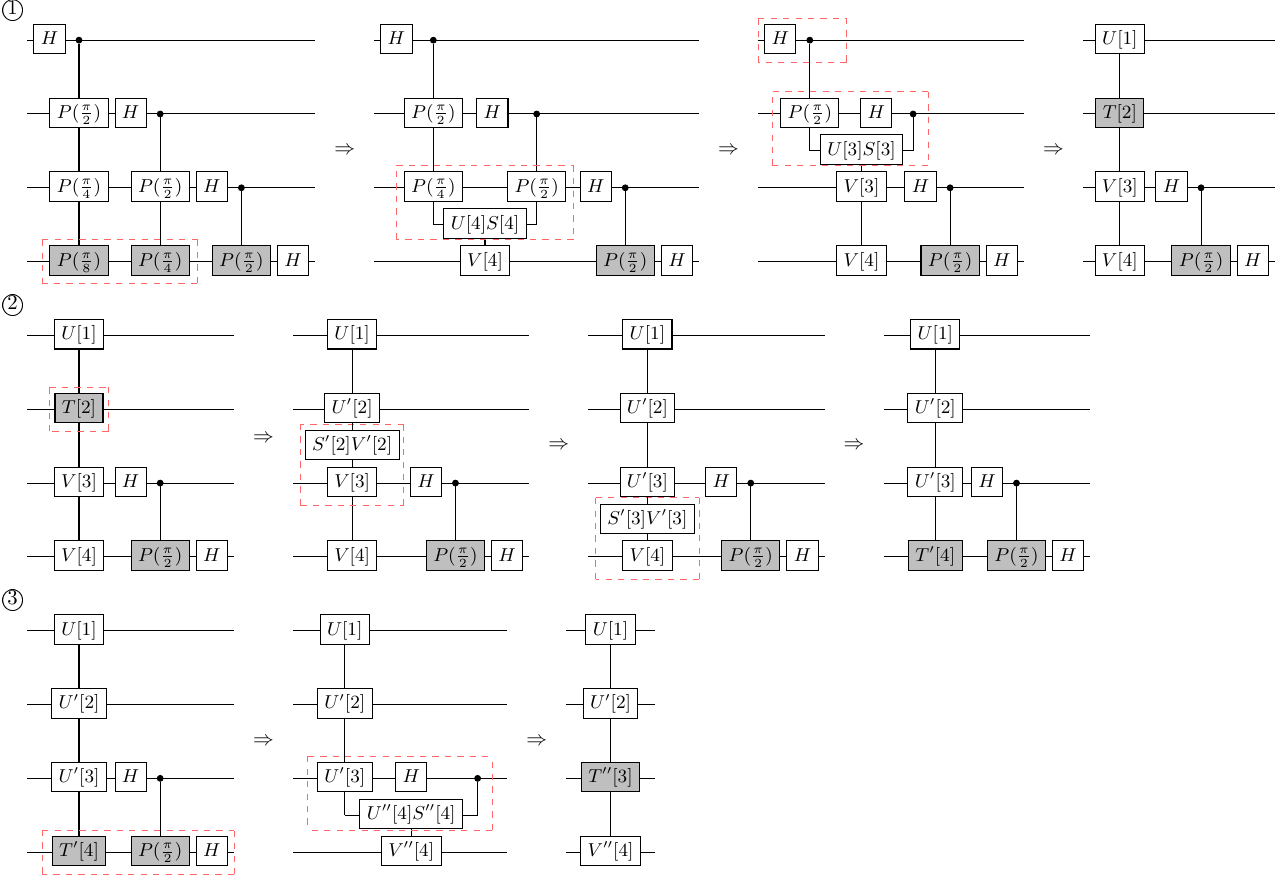}
    \caption{The diagrammatic illustration of contracting a 4-qubit QFT-TN into a QFT-MPO. In step \textcircled{1}, we contract the first two phase MPOs by doing a series of SVDs and we truncate small singular values or keep $\chi$ of them. Since site one is isometric already by contracting the $H$ tensor and the copy tensor, we stop at site 2, thus the orthogonality center at the end is at site 2. In step \textcircled{2}, we do a series of SVDs to push the orthogonality center to the last site, and again one truncates or keeps $\chi$ singular values. In step \textcircled{3}, we repeat the contraction for our current MPO and the next phase MPO. During the steps, we also contract any $H$ tensors encountered. The final MPO has the orthogonality center at site 3.}
    \label{fig:contract_QFT_TN}
\end{figure*}

We now show how to construct the QFT-MPO from the QFT-TN. The QFT-MPO is essentially a partially contracted tensor network of the QFT-TN. As the QFT-TN consists of several phase MPOs with trivial $H$ gates sandwiched between them, one direct method is to use the zip-up algorithm \cite{Stoudenmire_2010} for MPO contractions, which scales as $O(\chi^3 n^2)$ where $\chi$ is the largest bond dimension of the final QFT-MPO. The accuracy is guaranteed because all intermediate MPOs are well approximated by bond-dimension $\chi$; see Appendix~\ref{appendix:intermediate_MPOs} for details. The constant associated with the scaling is very small: constructing a 50-qubit QFT-MPO only takes around 1 second on a laptop with the ITensor software \cite{ITensor}, 

The zip-up algorithm works as follows: for two adjacent phase MPOs, we contract the orthogonality center of each MPO, which is located at the last site. Then we perform a singular value decomposition (SVD) of the resulting tensor, truncating small singular values up to a threshold or keeping only the first $\chi$ singular values, and leave the isometric tensor $V$ on the site while pushing $U$ and $S$ to the above site for contraction. The steps repeat until two phase MPOs merge into one. Since the phase MPOs have sizes no larger than $n$, the orthogonality center at this point will be moved to the first site or some site in the middle, thus to make it aligned with the next phase MPO's orthogonality center, we do a series of truncated SVDs to push it back into the last site. The process repeats for each pair of MPOs until all MPOs are contracted into one. During the steps, the $H$ tensors can also be contracted when convenient. A diagrammatic illustration is shown in Fig.~\ref{fig:contract_QFT_TN}. We conjectured that for any intermediate MPO, its last site combined with the next phase MPO's last site form an orthogonality center even \emph{before} the contraction, thus the zip-up algorithm for phase MPOs is completely stable; see Appendix~\ref{appendix:zip_up_stable} for detailed explanations. Therefore, the zip-up algorithm is an optimal algorithm for contracting the QFT-TN. 

In conclusion, by doing the zip-up algorithm over the QFT-TN we can construct a QFT-MPO with maximum $\chi$ bond dimension in $O(\chi^3 n^2)$ time. In principle, one could derive a closed-form expression for the QFT-MPO, since the QFT has a well-defined formula and the QFT-MPO's site tensors approach some fixed form as $n$ grows (due to the fact P-DPSSs approach DPSSs in the limit). One potential approach is to use the tensor train (MPS) cross-interpolation \cite{Savostyanov_2014}, a technique to approximate the MPS/MPO decomposition when the SVD is incalculable. Therefore, an $O(\chi^3 n)$ or even $O(\chi^2 n)$ algorithm might exist for constructing the QFT-MPO. However, we note that once a QFT-MPO is constructed it can be stored in a database and reused later, thus the construction time needs not to be not considered in algorithms. Additionally, our construction scheme has a very small constant associated with it, and is fast enough for the number of qubits required for most tasks. Therefore, improving the construction time of the QFT-MPO is less important compared to other parts of the algorithms employing the QFT-MPO. Nevertheless, it is interesting to see whether a closed-form expression exists for the QFT-MPO.

\subsection{Error for compressing the QFT-MPO}
\label{subsection:compression}

In this subsection, we analyze the error of compressing the QFT-MPO. In particular, we showed that both the average error and worst-case error are independent of $n$. The former is easy to benchmark with numerics and the latter guarantees the error of truncating a QFT-MPO will not be amplified by any input states.

Define $Q_n'$ to be the truncated MPO representation of $Q_n$ with bond-dimension $\chi$, then a convenient choice of the average error $e_{\text{avg}}$ is the average squared-norm of $Q_n - Q_n'$ acting on Haar random states:
\begin{equation}
    e_{\text{avg}} = \mathop{\mathbb{E}}_{\psi \sim \mu_H} \left[ \left\| (Q_n - Q_n') |\psi\rangle \right\|^2 \right]
\end{equation}
which reduces to
\begin{equation}
    e_{\text{avg}} = \frac{1}{2^n} \left\| Q_n - Q_n' \right\|_F^2
\end{equation}
from Haar integral calculations. We will show in Appendix~\ref{appendix:error_bounds} that $e_{\text{avg}}$ is bounded by $O(n e^{-\chi \log(\chi/3)}/\sqrt{\chi})$. Therefore $\chi$ needs to stay between constant and logarithmic growth to keep $e_{\text{avg}}$ from growing, and in practice, it is much closer to constant than logarithm.

We now demonstrate the scaling of $e_{\text{avg}}$ with some simple numerics. When restricted to $\chi=4$, the QFT-MPO with $n=7,8,9,10$ have $e_\text{avg}$ being $2.3 \times 10^{-6}, 4.2 \times 10^{-6}, 6.3 \times 10^{-6}, 8.3 \times 10^{-6}$, which is approximately linear; when restricted to $n=9$, the QFT-MPO with $\chi=5,6,7,8$ have errors $3.1 \times 10^{-8}, 9.4 \times 10^{-10}, 2.0 \times 10^{-13}, 3.2 \times 10^{-16}$, which decays at least exponentially.

While the average error is small, one cannot guarantee every input state is well-behaved. Therefore, a more powerful measure is the worst-case error, which is naturally defined as
\begin{equation}
    e_{\text{max}} = \sup_{|\psi\rangle:\langle\psi|\psi\rangle=1} \| (Q_n - Q_n')|\psi\rangle \|^2
\end{equation}
We will show in Appendix~\ref{appendix:error_bounds} that $e_{\text{max}}$ is also bounded by $O(n e^{-\chi \log(\chi/3)}/\sqrt{\chi})$, the same order as $e_{\text{avg}}$. Therefore one can safely apply $Q_n'$ to any state without introducing a large error.

\section{Application: QFT versus the FFT}
\label{section:QFT_vs_FFT}
The QFT was originally intended to be run on a quantum computer, taking
advantage of a quantum computer's ability to manipulate a state of size $2^n$ using only $n$ qubits.

The observation that the QFT can be realized as a highly compressed MPO suggests an alternate view: the QFT
can be used as a \emph{quantum-inspired classical algorithm} for computing a discrete Fourier transform. 
Here the role of the quantum computer is played by an MPS tensor network representing
the function to be transformed. Much like a quantum computer, MPS can achieve exponential
compression by storing functions defined on $2^n$ grid points using only $n$ tensors. The key limitation of the MPS
format is that mainly functions that are smooth or have a self-similar structure can be compressed \cite{Khoromskij_2011,Ryzhakov_2022}, though we will demonstrate a function with sharp cusps which also compresses well.

The QFT as a classical algorithm has already been proposed and was dubbed the \emph{superfast Fourier transform}, 
since it was observed to outperform the fast Fourier transform (FFT) in certain cases \cite{Dolgov_2012,Garcia-Ripoll}. 
However, the observation that the QFT itself could also be represented as a low bond dimension tensor network has not been used
in previous work on this algorithm.

\begin{figure*}[th]
    \centering
    \includegraphics[width=1.5\columnwidth]{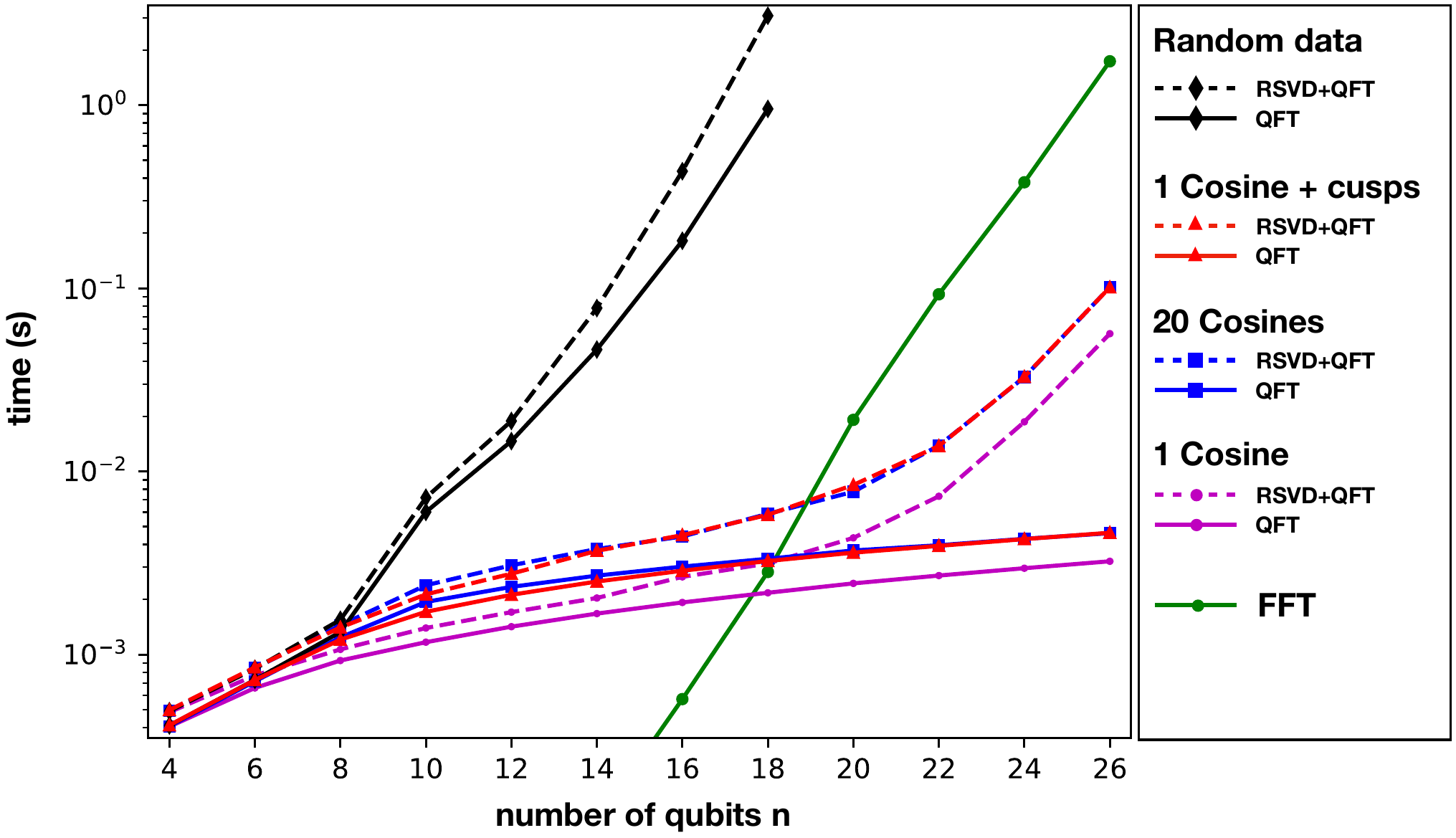}
    \caption{Time in seconds to compute the discrete Fourier transform using either the fast Fourier transform (FFT) algorithm for data of size $2^n$ or the quantum Fourier transform (QFT) MPO acting on various functions represented as MPS with $n$ sites (qubits). Dashed curves include the time to convert the data to the MPS format using a randomized SVD (RSVD) algorithm plus the application of the QFT, while solid curves are the time of the QFT step only. Timings were performed on a 2021 10-core Macbook Pro with an M1 Max processor, using the FFTW and ITensor software \cite{ITensor}.
    \label{fig:qft_timings}}
\end{figure*}

Here we revisit the idea, taking advantage of the MPO form of the QFT to perform the superfast Fourier transform and 
comparing its performance to the FFT. Depending on the problem and context, the function being
transformed may already be represented as an MPS, or may need to be converted to an MPS before the QFT can be applied. 
We therefore perform two sets of timings: one of just the application of the QFT and the other including
the time to convert the function to an MPS. 

We require the length of the data vector to be a power of two, and the data is reshaped into the form of an n-qubit tensor, with each index corresponding to a binary digit.  Direct compression of the tensor into MPS form would involve the singular value decomposition, which in the worst case involves $\sqrt{N}\times\sqrt{N}$ matrices where $N=2^{n}$, and taking a time proportional to $N^{3/2}=2^{3 n/2}$. Assuming the data is compressible, we instead utilize randomized singular value decompositions \cite{Halko}, a fairly simple and very robust method, to compress more efficiently. For an MPS with maximum bond dimension $\chi_m$, the compression time for the central bond becomes $O(\chi_m N + \chi_m^2\sqrt{N})$.  Here the dominant $\chi_m N$ term consists of a matrix-matrix multiply, one of the most efficient linear algebra operations, of the $\sqrt{N}\times\sqrt{N}$ data matrix times a $\sqrt{N}\times\chi_m'$ matrix of random numbers, where $\chi_m'-\chi_m\sim$ 5 to 10. The subdominant term corresponds to a subsequent standard SVD on the product $\sqrt{N}\times\chi_m'$ matrix. Moving from the center to both sides of the MPS, one decomposes matrices with sizes $\sqrt{N}/2 \times 2\chi_m$,  $\sqrt{N}/4 \times 2\chi_m...$, etc.,
which contribute to the prefactor of the subdominant term only.  The overall compression time for the entire MPS is thus $O(\chi_m N + \chi_m^2 \sqrt{N})$. For constant $\chi_m$, we have the complexity $O(N) = O(2^n)$. This is to be compared with the $O(N\log N) = O(2^n n)$ time of a standard FFT for the same size of data.

The timings we obtain are shown in Fig.~\ref{fig:qft_timings}. In the background of the plot, we show the time taken by the FFT as a function of $n$ for data of length $2^n$. Note that the FFT operates as a ``black box'' insensitive to the data being transformed, whereas the QFT approach is only fast if the data compresses well into an MPS of small or moderate bond dimension. We determine the MPS bond dimension adaptively by setting a truncation error threshold of $10^{-12}$ in the randomized SVD compression procedure. The construction of the QFT MPO is done in advance and not included, but its construction time of $O(n^2)$ is negligible anyway. To apply the QFT MPO to the data MPS we use the ``fitting'' approach based on a single-site, DMRG style algorithm \cite{Stoudenmire_2010,Verstraete_2004}.

Using the QFT approach on random data is much slower than the FFT, as shown in Fig.~\ref{fig:qft_timings}. This is because the MPS bond dimensions saturate their maximum values for this case, which are powers of two up to the central bond having a dimension of $2^{n/2}$. Therefore the QFT and MPS approach gives no advantage for random data.

Using structured data, we find a crossover point around $n=18$ where the QFT approach is \emph{faster} than the FFT, even accounting for the time needed to convert the data to MPS form. We consider three example functions, with more details about these functions given in Appendix~\ref{appendix:functions}. The first is 
a single cosine function making one period over the interval $[0,1]$. 
This function compresses exactly into an MPS of bond dimension $\chi_m=2$.
The next example is a sum of twenty cosines of various frequencies, combining together to make a function
resembling a $\text{sinc}$ function centered in the interval $[0,1]$. The last example is a single cosine plus a sum of four functions of the form $a e^{-b |x-c|}$ for various values of $a$, $b$, and $c$, which results in the function having sharp cusps. For these last two examples, the MPS bond dimensions are around $\chi_m \sim 10$ with the dimensions being somewhat more uniform for the cosine-plus-cusps example. The time of the QFT step is approximately linear in $n$, though it depends on the specific bond dimensions needed to represent a function on a grid size of $2^n$, which can have a more complicated but mild dependence on $n$.

Beyond $n>18$ the QFT approach becomes orders of magnitude faster than the FFT if one does not include the cost of constructing the data MPS. This is reasonable in certain contexts, such as if the function to be transformed is a solution of a differential equation carried out using the MPS format  \cite{Khoromskij2009, Lubasch, Gourianov2022} or the result of a quantum simulation with an MPS state representation \cite{Daley,White2004,Vidal2007}. Even including the construction of the data MPS, with data as compressible as in these simple cases one finds about one order of magnitude improvement over the FFT. Efficient algorithms for compressing functions into MPS are an ongoing area of research \cite{Oseledets2010tt,Dolgov2020,Ryzhakov_2022,Nunez_Fernandez}.

\section{Conclusion}
\label{section:conclusion}
We have shown that the maximum-entanglement property of the standard QFT mainly comes from the trivial bit-reversal operation, while the core part of QFT has Schmidt coefficients decaying exponentially quickly. We have given rigorous bounds on the Schmidt coefficients, demonstrating the exponential decay,
and we have given a method to construct the constant-bond-dimension QFT-MPO with an exponentially small error in $O(n^2)$ time using a tensor network representation of QFT. Thus simulating the QFT on a classical computer with high precision can be done in $O(n)$ time if a low-bond-dimension MPS representation of the initial state is given. In tests comparing the QFT and FFT for Fourier transforming data in a purely classical context, we have found that when the data is compressible, the QFT approach has better scaling and can be significantly faster, even when accounting for the time needed to convert the data to an MPS. We emphasize that this does not mean QFT is efficiently classically simulable in general (e.g. states in Shor's algorithm), since for general states one expects exponentially large MPS, thus both conversion and applying the QFT-MPO take exponential time.

The low-entanglement property suggests that the bit-reversed QFT might be a more natural way to define the QFT, since we introduced a lot of unnecessary entanglement from forcing the output bits to maintain the ordering. An interesting follow-up question is if the entanglement structure of the QFT over more general classes of finite groups has similar properties. One may also wonder whether the low-entanglement property can lead to better methods for implementing the QFT on a quantum computer, such as designing circuits with a shorter depth and fewer ancillas, or employing the QFT's underlying Hamiltonian dynamics. From the quantum information point of view, the QFT's low-entanglement might have many implications, and we leave it as an open question to identify them.

On the other hand, in the classical setting, the biggest open question is what classes of functions can take advantage of the MPS representation to allow the QFT-MPO method to outperform the FFT. At the same time, finding better methods for converting the data into the MPS is also crucial. Since conversion dominates the computation time in the QFT-MPO algorithm, it is also important to classify cases where conversion is not needed, such as algorithms where a tensor network structure was given in the first place. It is also interesting to ask if this result can be connected to the dequantization framework \cite{Tang_2022, Tang_2019, Chia_2020, Gharibian_2022}, since the MPS and the QFT-MPO have intrinsic efficient sampling and query accesses. More importantly, if compressible functions frequently appear in practical tasks, such as condensed matter problems, image processing, and medical imaging, etc., one may expect a wide application of the QFT-MPO method.

\begin{acknowledgements}
We thank Artem Strashko for helpful discussions regarding prior work on approximations and simulations of the QFT. SRW was
supported by the NSF through grant DMR-2110041.
\end{acknowledgements}

\clearpage

\begin{appendices}

\section{Proof of the QFT decomposition}
\label{appendix:qft_decomposition}
In this appendix, we prove that the decomposition
\begin{equation}
    Q_n = \left(I_j \otimes Q_{n-j}\right) \Omega_{n,j} \left(Q_{j} \otimes I_{n-j}\right)
    \label{eq:rQFT_decomp_app}
\end{equation}
termed as the generalized QFT circuit \cite{Cleve_00}, holds for any $n$ and $j$. Recall that $Q_n$ is the quantum Fourier transform with the order of output bits reversed
\begin{multline}
    Q_n|q_1 q_2 ... q_n\rangle = \frac{1}{2^{n/2}} (|0\rangle + e^{2\pi i 0.q_1 q_2...q_n}|1\rangle) ~ \otimes \\
    (|0\rangle + e^{2\pi i 0.q_{2}...q_n}|1\rangle) \otimes... (|0\rangle + e^{2\pi i 0.q_n}|1\rangle)
    \label{eq:rQFT_def_app}
\end{multline}
and $\Omega_{n,j}$ is the diagonal unitary composed of controlled-phase gates $P_{l,m}(\theta)$, i.e.
\begin{equation}
    \Omega_{n,j}=\prod_{l=1}^j \prod_{m=j+1}^{n} P_{l,m}(\frac{\pi}{2^{m-l}})
\end{equation}
We will prove Eq.~\eqref{eq:rQFT_decomp_app} by tracking the states after sequentially applying $Q_{j} \otimes I_{n-j}$, $\Omega_{n,j}$, $I_j \otimes Q_{n-j}$ on an arbitrary product state $|q_1 q_2 ... q_n\rangle$, and show the final state matches Eq.~\eqref{eq:rQFT_def_app}. We start by considering the linear map $Q_j \otimes I_{n-j}$, which is the smaller bit-reversed QFT applied on the first $j$ qubits:
\begin{multline}
    (Q_j \otimes I_{n-j})|q_1 q_2 ... q_n\rangle = \\ \frac{1}{2^{j/2}} (|0\rangle + e^{2\pi i 0.q_1 q_2...q_j}|1\rangle) \otimes... \\
    (|0\rangle + e^{2\pi i 0.q_j}|1\rangle) \otimes |q_{j+1}\rangle \otimes... |q_{n}\rangle
\end{multline}
We then consider the effect of applying $\Omega_{n,j}$. Recall that in the subspace spanned by qubits $l$ and $m$, $P_{l,m}(\theta)$ can be represented as the linear map
\begin{equation}
    P_{l,m}(\theta)|q_lq_m\rangle = e^{i \theta q_lq_m} |q_lq_m\rangle
\end{equation}
Therefore, by combining phases and attaching them to each of qubits $1$ to $j$, the $\Omega_{n,j}$ corresponds to a linear map 
\begin{multline}
    \Omega_{n,j}|q_1 q_2 ... q_n\rangle = \left(\exp(i\pi q_1 \sum_{m=j+1}^n \frac{q_m}{2^{m-1}})|q_1\rangle\right) \otimes... \\ \left(\exp(i\pi q_j \sum_{m=j+1}^n \frac{q_m}{2^{m-j}})|q_j\rangle\right) \otimes |q_{j+1}\rangle \otimes... |q_{n}\rangle
\end{multline}
In terms of the shorthand notation $0.q_1q_2...q_n$, we can express each phase on qubit $l$ as
\begin{equation}
\begin{split}
    & \exp(i\pi q_l \sum_{m=j+1}^n \frac{q_m}{2^{m-l}})\\
    = & \exp(i2\pi \sum_{m=j+1}^n \frac{q_m}{2^{m-l+1}})^{q_{l}} \\
    = & \exp(i2\pi 0.\underbrace{0...0}_{j-l+1}q_{j+1}...q_{n})^{q_{l}}
\end{split}
\end{equation}
Therefore, applying $\Omega_{n,j}$ to the state $(Q_j \otimes I_{n-j})|q_1q_2....q_n\rangle$ simply fills the phase of the $|1\rangle$ states of the first $j$ qubits, and the overall linear map matches $Q_n$ for the first $j$ qubits:
\begin{multline}
    \Omega_{n,j}(Q_j \otimes I_{n-j})|q_1 q_2 ... q_n\rangle = \\ \frac{1}{2^{j/2}} (|0\rangle + e^{2\pi i 0.q_1 q_2...q_jq_{j+1}...q_n}|1\rangle) \otimes... \\
    (|0\rangle + e^{2\pi i 0.q_jq_{j+1}...q_n}|1\rangle) \otimes |q_{j+1}\rangle \otimes... |q_{n}\rangle
\end{multline}
The final step $\left(I_j \otimes Q_{n-j}\right)$ is trivial, since it does not affect the first $j$ qubits and acts as a smaller bit-reversed QFT on the last $n-j$ qubits. One can then verify both sides of Eq.~\ref{eq:rQFT_decomp_app} achieve the same transformation.

\section{Spectral Concentration}
\label{appendix:spectral_concentration}
In this appendix, we introduce the definition of the \emph{spectral concentration} problems, which were originally studied by D. Slepian, H. Landau, and H. Pollack \cite{Slepian_1961, Landau_1961, Landau_1962, Slepian_1964, Slepian_1978}. Consider the \emph{discrete-time Fourier transform} (DTFT) of a finite series $v = \{v_x\}$ with $x = 1, 2, 3, ... N'$,
\begin{equation}
    F_{DT}(v)( \tilde{t})=\sum _{x=1}^{N'} v_x e^{-2\pi i x \tilde{t}}
    \label{eq:DTFT}
\end{equation}
where $\tilde{t}$ is a continuous variable. Notice because the sampling interval $\Delta x = 1$, DTFT has period 1, i.e. $F_{DT}(\tilde{t}) = F_{DT}(\tilde{t}+1)$, therefore we can restrain the domain of $\tilde{t}$ to $[-1/2, 1/2]$. The spectral concentration problem then asks the following: for a given frequency $W$ such that $0 < W < 1/2$, find the finite series $v=\{v_x\}$ that maximizes the spectral concentration quantity
\begin{equation}
    \tilde{\lambda}_{N',W}(v) = \frac{\int_{-W}^W ||F_{DT}(v)(\tilde{t})||^2 d\tilde{t}}{\int_{-1/2}^{1/2} ||F_{DT}(v)(\tilde{t})||^2 d\tilde{t}}
    \label{eq:spectral_concentration}
\end{equation}
which can be interpreted as the ratio of power of $F_{DT}(\tilde{t})$ contained in the frequency band $[-W,W]$ to the power of $F_{DT}(\tilde{t})$ contained in the entire frequency band $[-1/2,1/2]$. By substituting Eq.~\eqref{eq:DTFT} to Eq.~\eqref{eq:spectral_concentration}, we get the expression
\begin{equation}
    \begin{split} 
    \tilde{\lambda}_{N',W}(v)
    & = \frac{\int_{-W}^W ||F_{DT}(v)(\tilde{t})||^2 d\tilde{t}}{\int_{-1/2}^{1/2} ||F_{DT}(v)(\tilde{t})||^2 d\tilde{t}}\\
    & = \frac{\int_{-W}^W \sum_{x=1}^{N'} \sum_{y=1}^{N'} v_{x}^* v_{y} e^{2\pi i \tilde{t} (x-y)} d\tilde{t}}{\sum_{x=1}^{N'} ||v_x||^2}\\
    & = \sum_{x=1}^{N'} v_x^* \sum_{y=1}^{N'} \frac{\sin(2\pi W (x-y))}{\pi (x-y)}v_y  \Big/ \sum_{x=1}^{N'} ||v_x||^2
    \end{split}
\end{equation}
and thus finding the maximum spectral concentration corresponds to solving the eigenvalue equation
\begin{equation}
    \sum_{y=1}^{N'} \frac{\sin(2\pi W (x-y))}{\pi (x-y)} \tilde{v}_{N',W}^{k,y} = \tilde{\lambda}_{N',W}^{k} \tilde{v}_{N',W}^{k,x}
    \label{eq:app_DPSS_eig_eq}
\end{equation}
The eigenvectors $\{\tilde{v}_{N',W}^{k}\}$ are called discrete prolate spheroidal sequences (DPSS) \cite{Slepian_1978} and eigenvalues $\{\tilde{\lambda}_{N',W}^{k}\}$ are spectral quantities of DPSSs as input sequences, i.e. $ \tilde{\lambda}_{N',W}^{k} = \tilde{\lambda}_{N',W}(\tilde{v}_{N',W}^{k})$. The largest eigenvalue $\tilde{\lambda}_{N',W}^{0}$ is the solution to the spectral concentration problem, i.e. the maximum spectral concentration quantity one can get. The following eigenvalues $\tilde{\lambda}_{N',W}^{k}$ for $k \ge 1$ have the meaning that, in the subspace orthogonal to the first $k$ DPSSs, $\tilde{\lambda}_{N',W}^{k}$ is the maximum spectral concentration one can get among all vectors in the subspace.

We can also define the discrete version of this problem. Consider the truncated \emph{discrete Fourier transform} (DFT) instead of the DTFT on the same sequence $v = \{v_x\}$,
\begin{equation}
    F_{D}(v)_t = \sum_{x=1}^{N'} v_{x} e^{-2\pi i x t/N}
    \label{eq:DFT}
\end{equation}
where by ``truncated'' we mean we only sum up to $N' \leq N$. The DFT is identical to the DTFT except that the continuous variable $\tilde{t} \in [-1/2, 1/2]$ is changed into a discrete sequence $\{t/N\}$ with $t=-N/2+1/2, -N/2+3/2, ... N/2-1/2$, and $W$ is a number in $[0, 1/2]$ so that $2NW$ is an integer. Note that due to the convention, the operator $F_{n,j}$ appears in Section~\ref{subsection:QFT_LE_proof} has opposite phases and rows corresponding to $x$ instead of columns, thus the truncated DFT with $N=2^n$, $N'=2^j$ and $2W=1/2^j$ corresponds to the operator $F_{n,j}^\dagger$ instead of $F_{n,j}$. Similarly, we define the quantity $\lambda_{N,N',W}$ and expand the expression
\begin{equation}
    \begin{split} 
    \lambda_{N,N',W}(v)
    & = \frac{\sum_{t=-NW+1/2}^{NW-1/2} ||F_D(v)_t||^2}{\sum_{t=-N/2+1/2}^{N/2-1/2} ||F_D(v)_t||^2}\\
    & = \frac{\sum_{t=-NW+1/2}^{NW-1/2} \sum_{x=1}^{N'} \sum_{y=1}^{N'} v_{x}^* v_{y} e^{2\pi i \frac{t}{N} (x-y)}}{N\sum_{x=1}^{N'} ||v_x||^2}\\
    & = \sum_{x=1}^{N'} v_x^* \frac{1}{N}\sum_{y=1}^{N'} \frac{\sin(2\pi W (x-y))}{\sin(\pi (x-y)/N)}v_{y}   /\sum_{x=1}^{N'} ||v_x||^2
    \end{split}
\end{equation}
where the last line utilizes the sum of geometric series. The eigenvalue equation corresponding to the discrete version is thus
\begin{equation}
    \frac{1}{N}\sum_{y=1}^{N'} \frac{\sin(2\pi W (x-y))}{\sin(\pi (x-y)/N)}v_{N,N',W}^{k,y} = \lambda_{N,N',W}^{k} v_{N,N',W}^{k,x}
\end{equation}
whose eigenvalues are called periodic discrete prolate spheroidal sequences (P-DPSSs) \cite{Xu_1984}. As $N \rightarrow \infty$, the DFT approaches the DTFT, and thus so do the eigenvalue equations. Since the eigenvalue equations are identical, solving the discrete version of the spectral concentration problem gives the Schmidt coefficients of the QFT operator $Q_n$. 

In the original spectral concentration problem, we can also consider the inverse DTFT that maps continuous frequency functions into a sequence of samples,
\begin{equation}
    F_{DT}^{-1}(f)_x = \int_{-W}^{W} f(\tilde{t}) e^{2\pi i x \tilde{t}} ~ d\tilde{t}
\end{equation}
To avoid extra phases arising from the time-shifting theorem of the Fourier transform, from now on we shift $x$ from $1,...N'-1$ to $-N'/2+1/2,...N'/2-1/2$ for both $F_{DT}^{-1}$ and $F_{DT}$. The DPSS eigenvalue equation becomes
\begin{equation}
    \begin{split} 
    F_{DT}^{-1}\left(F_{DT}\left(v_{N',W}^{k}\right)\right) = \tilde{\lambda}_{N',W}^{k} v_{N',W}^{k}
    \end{split}
\end{equation}
which can be turned into Eq.~\eqref{eq:app_DPSS_eig_eq} by contracting $F_{DT}^{-1}$ and $F_{DT}$ first (the shift of $x$ and $y$ does not change the equation since the operator's elements only depend on $x-y$). Because $F_{DT}$ is a compact operator, $F_{DT}^{-1}\left(F_{DT}\left(\cdot \right)\right)$ and $F_{DT}\left(F_{DT}^{-1}\left(\cdot \right)\right)$ will share the same non-zero eiegnavlues. This allows us to instead consider the equation
\begin{equation}
    \begin{split} 
    F_{DT}\left(F_{DT}^{-1}\left( f_{N',W}^{k} \right)\right) = \tilde{\lambda}_{N',W}^{k} f_{N',W}^{k}
    \end{split}
\end{equation}
where $f_{N',W}^{k}$ are continous functions known as the \emph{discrete prolate spheroidal wave functions} (DPSWFs) \cite{Slepian_1978}. Contracting the operators $F_{DT}\left(F_{DT}^{-1}\left(\cdot \right)\right)$ gives the integral equation
\begin{equation}
    \begin{split}
    \sum_{x=-N'/2+1/2}^{N'/2-1/2} \int_{-W}^{W} e^{2\pi i x (\tilde{t} - \tilde{s})} f_{N',W}^{k}(\tilde{s})~d\tilde{s} &= \tilde{\lambda}_{N',W}^{k} f_{N',W}^{k}(\tilde{t})\\
    \int_{-W}^{W} \frac{\sin(N'\pi (\tilde{t}-\tilde{s}))}{\sin(\pi (\tilde{t}-\tilde{s}))} f_{N',W}^{k}(\tilde{s}) ~ d\tilde{s} &= \tilde{\lambda}_{N',W}^{k} f_{N',W}^{k}(\tilde{t})
    \end{split}
\end{equation}
which we call the DPSWF eigenvalue equation.

For the case of DFT, $F_{D}^{-1}\left(F_{D}\left(\cdot \right)\right)$ and $F_{D}\left(F_{D}^{-1}\left(\cdot \right)\right)$ give the same type of eigenvalue equation except that variables $W, N', N$ are in different places. One can also consider the case of continuous Fourier transform, which instead gives the sinc kernel integral operator and whose eigenfunctions are known as \emph{prolate spheroidal wave functions} (PSWFs), which were the original problem D. Slepian, H. Landau and H. Pollack studied \cite{Slepian_1961, Landau_1961, Landau_1962}. 

\section{Upperbounds on the DPSSs' eigenvalues}
\label{appendix:DPSS_upperbounds}
In this appendix, we prove the exponential decay of the eigenvalues of the DPSSs. Adopting techniques from \cite{Bonami_2021, Boulsane_2020}, we will consider the DPSWF operator $F_{DT}\left(F_{DT}^{-1}\left( \cdot \right)\right)$ which has the same spectrum as the DPSS operator $F_{DT}^{-1}\left(F_{DT}\left( \cdot \right)\right)$.  For an introduction to DPSS and DPSWF, see Appendix~\ref{appendix:spectral_concentration}.

Recall that $F_{DT}\left(F_{DT}^{-1}\left( \cdot \right)\right)$ is a self-adjoint compact operator, therefore, its eigenvalues obey the Courant-Fischer-Weyl min-max principle
\begin{equation}
\begin{split}
    \tilde{\lambda}_{N',W}^{k} &= \min_{\mathcal{S}_k} \max_{f \in \mathcal{S}_k^\perp} \frac{\langle F_{DT}\left(F_{DT}^{-1}\left( f \right)\right), f \rangle}{\langle f, f \rangle} \\
    &= \min_{\mathcal{S}_k} \max_{f \in \mathcal{S}_k^\perp, \|f\|=1 } \| F_{DT}^{-1}(f) \|^2 \\
\end{split}
\end{equation}
where $\mathcal{S}_k$ is a $k$-dimensional subspace of the $L^2$ function space in the domain $[-W, W]$. The min-max principle tells us that we can specify a particular subspace and guarantee to have an upper bound on the eigenvalues $\{\tilde{\lambda}_{N',W}^{k}\}$. We consider the case that
\begin{equation}
    \mathcal{S}_k = \text{Span}\left\{\hat{P}_0, \hat{P}_1, ... \hat{P}_{k-1} \right\}
\end{equation}
where $\hat{P}_l$ are normalized functions of the form
\begin{equation}
    \hat{P}_l(\tilde{t}) = \sqrt{\frac{2l+1}{2W}} P_l\left(\frac{\tilde{t}}{W}\right)
\end{equation}
where $P_l$ are the Legendre polynomials of order $l$, and the extra normalization constant comes from the fact that
\begin{equation}
    \left\| P_l\left(\frac{\tilde{t}}{W}\right) \right\|^2 
    = W \int_{-1}^{1}  \left|P_l\left(\tilde{s}\right) \right|^2 ~ d\tilde{s}
    = \frac{2W}{2l+1}
\end{equation}
The normalized functions in the subspace $\mathcal{S}_k^\perp$ can thus be expanded as
\begin{equation}
    f(\tilde{t}) = \sum_{l = k}^\infty a_l \hat{P}_l(\tilde{t}), ~~ \sum_{l = k}^\infty |a_l|^2 = 1
\end{equation}
and by the linearity of the DTFT, we have
\begin{equation}
    F_{DT}^{-1}\left(f(\tilde{t})\right)_x = \sum_{l = k}^\infty a_l F_{DT}^{-1}\left(\hat{P}_l(\tilde{t})\right)_x
    \label{eq:app_FT_f}
\end{equation}
For each term, we can re-express in terms of the Fourier transform of Legendre polynomials evaluated in the interval $[-1, 1]$,
\begin{equation}
\begin{split}
    F_{DT}^{-1}\left(\hat{P}_l(\tilde{t})\right)_x
    &= \sqrt{\frac{2l+1}{2W}} \int_{-W}^{W}  P_l\left(\frac{\tilde{t}}{W}\right) e^{-2\pi i x \tilde{t}} ~ d\tilde{t} \\
    &= \sqrt{\frac{W(2l+1)}{2}} \int_{-1}^{1}  P_l\left(\tilde{s}\right) e^{-2\pi i W x \tilde{s}} ~ d\tilde{s} \\
\end{split}
\label{eq:FT_Pl}
\end{equation}
whose analytical solutions were known, e.g. see \cite{Olver_2010}:
\begin{equation}
    \int_{-1}^{1}  P_l\left(\tilde{y}\right) e^{i \tilde{x}\tilde{y}} ~ d\tilde{y} = i^l \sqrt{\frac{2\pi}{\tilde{x}}} J_{l+\frac{1}{2}}(\tilde{x})
\end{equation}
where $J_{l+1/2}$ are half-order Bessel functions. Moreover, it is known that Bessel functions have the fast decay
\begin{equation}
    |J_\alpha (\tilde{z})| \leq \frac{\left|\frac{\tilde{z}}{2}\right|^\alpha}{\Gamma(\alpha+1)}
\end{equation}
where $\Gamma$ is the Gamma function, which is lower-bounded by
\begin{equation}
    \Gamma(\alpha+1) \ge \sqrt{2e}\left(\frac{\alpha+\frac{1}{2}}{e} \right)^{\alpha+\frac{1}{2}}
\end{equation}
Thus we can obtain the upper bound
\begin{equation}
\begin{split}
    \left| \int_{-1}^{1} P_l\left(\tilde{s}\right) e^{-2\pi i W x \tilde{s}} ~ d\tilde{s} \right|^2 &\le \left| i^l \sqrt{\frac{2\pi}{-2\pi W x}} J_{l+\frac{1}{2}}(-2\pi W x) \right|^2\\
    &\le \frac{1}{W} \frac{\left(\pi W \right)^{2l+1}}{2e\left(\frac{l+1}{e} \right)^{2l+2}} |x|^{2l}\\ &= \frac{1}{2(l+1) W}\left(\frac{e\pi W}{l+1}\right)^{2l+1} |x|^{2l}\\
\end{split}
\label{eq:app_FT_Pl_bound}
\end{equation}
We also note that summing over $|x|^{2l}$ is upper-bounded by the corresponding integral
\begin{equation}
    \sum_{x=-(N'-1)/2}^{(N'-1)/2} |x|^{2l} \leq \int_{-N'/2}^{N'/2} |\tilde{x}|^{2l} d\tilde{x} = \frac{2}{2l+1}\left(\frac{N'}{2}\right)^{2l+1}
\end{equation}
Therefore, summing over $x$ in Eq.~\eqref{eq:app_FT_Pl_bound} gives
\begin{multline}
    \sum_{x=-(N'-1)/2}^{(N'-1)/2} \left| \int_{-1}^{1} P_l\left(\tilde{s}\right) e^{-2\pi i W x \tilde{s}} ~ d\tilde{s} \right|^2 \\
    \leq \frac{1}{(2l+1)(l+1)W}\left(\frac{e\pi N' W}{2(l+1)}\right)^{2l+1}
    \label{eq:app_FT_Pl_2norm}
\end{multline}
Combing this equation and Eq.~\eqref{eq:FT_Pl}, we obtain the upper bound
\begin{equation}
\begin{split}
    \left\|F_{DT}^{-1}\left(\hat{P}_l(\tilde{t})\right)_x\right\|^2 &= \sum_{x=-(N'-1)/2}^{(N'-1)/2} \left| F_{DT}^{-1}\left(\hat{P}_l(\tilde{t})\right)_x \right|^2 \\
    &\leq \frac{1}{2(l+1)}\left(\frac{e\pi N' W}{2(l+1)}\right)^{2l+1}
\end{split}
\end{equation}
Then by applying the triangle inequality and the Cauchy-Schwarz inequality, one can get the following upper bound for the squared norm of Eq.~\eqref{eq:app_FT_f},
\begin{equation}
\begin{split}
    \| F_{DT}^{-1}\left(f(\tilde{t})\right)_x \|^2 &\leq \left( \sum_{l = k}^\infty |a_l| \left\|F_{DT}^{-1}\left(\hat{P}_l(\tilde{t})\right)_x\right\| \right)^2 \\
    &\leq \left(\sum_{l = k}^\infty |a_l|^2\right) \left(\sum_{l = k}^\infty \left\|F_{DT}^{-1}\left(\hat{P}_l(\tilde{t})\right)_x\right\|^2\right) \\
    &= \sum_{l = k}^\infty \frac{1}{2(l+1)}\left(\frac{e\pi N' W}{2(l+1)}\right)^{2l+1} \\
    &\leq \frac{1}{2(k+1)}\sum_{l = k}^\infty \left(\frac{e\pi N' W}{2(l+1)}\right)^{2l+1}
\end{split}
\label{eq:app_FT_f_sum}
\end{equation}
where for the summation term, by replacing $l$ with $k$ and breaking it into two parts, one obtains
\begin{equation}
    \left(\frac{e\pi N' W}{2(l+1)}\right)^{2l+1} \leq \left(\frac{e\pi N' W}{2(k+1)}\right)^{2k+1} \left(\frac{e\pi N' W}{2(k+1)}\right)^{2l-2k} 
\end{equation}
The sum over the latter term is a simple sum of geometric series. If $k > e\pi N' W/2 - 1$, the sum converges and evaluates to
\begin{equation}
    \sum_{l=k}^\infty \left(\frac{e\pi N' W}{2(k+1)}\right)^{2l-2k} = \frac{4(k+1)^2}{4(k+1)^2 - (e\pi N' W)^2}
    \label{eq:app_DPSS_bound_constant}
\end{equation}
Therefore, by substituting back to Eq.~\eqref{eq:app_FT_f_sum} one concludes that for $k > e\pi N' W/2 - 1$,
\begin{equation}
    \begin{split}
    \tilde{\lambda}_{N',W}^{k} &\leq \| F_{DT}^{-1}\left(f(\tilde{t})\right)_x \|^2 \\ &\leq  \frac{2(k+1)}{4(k+1)^2 - (e\pi N' W)^2}
    \left(\frac{e\pi N' W}{2(k+1)}\right)^{2k+1}
    \end{split}
\end{equation}
For the particular case that $2N'W = 1$ and $k \ge 2$ (chosen as the first integer greater than $e\pi/4-1$), the leading term is always below $1/k$, obtaining that for $k \ge 2$,
\begin{equation}
    \tilde{\lambda}_{j}^{k} \leq \frac{1}{k} \exp \left(-(2k+1)\log\left(\frac{4k+4}{e\pi}\right)\right)
\end{equation}
where $\tilde{\lambda}_{j}^{k}$ are eigenvalues of the case when $N'=2^j$ and $2W = 1/2^j$.

\section{Eigenvalues of DPSSs and P-DPSSs}
\label{appendix:spectral_dynamics}
In this appendix, we prove that in the context of the QFT, with a mild assumption, the non-dominant eigenvalues of DPSSs are upper bounds to those of P-DPSSs. Consider the P-DPSS matrix arising from the QFT
\begin{equation}
    T_{n,j}[x, y] = \frac{1}{2^n} \frac{\text{sin}(\pi(x-y)/2^j)}{\text{sin}(\pi(x-y)/2^n)}
\end{equation}
where $[x,y]$ denotes the $(x+1)$-th row and $(y+1)$-th column and $x,y \in [0, 2^j-1]$. We argue that as $n$ increases, the first (largest) eigenvalue of $T_{n,j}$ will decrease and all other eigenvalues will increase. Therefore, if we consider matrices arising from the limit $n \rightarrow \infty$, the DPSS matrix arising from an infinite-sized QFT
\begin{equation}
    \tilde{T}_{j}[x, y] = \lim_{n \rightarrow \infty} T_{n,j}[x, y] = \frac{\text{sin}(\pi(x-y)/2^j)}{\pi(x-y)}
\end{equation}
any upper bound on the non-leading eigenvalues ($\tilde{\lambda}_{j}^k$) of $\tilde{T}_{j}$ can be directly applied to those of $T_{n,j}$ ($\lambda_{n,j}^k$) for all $n \ge 0$. 
 
The assumption we make is that both $T_{n,j}$ and $\tilde{T}_{j}$ have the set of eigenvalues for $k \ge 2NW = 1$ forming a smoothly decaying curve, which is evident from numerics and the early analysis of the spectral concentration problem \cite{Slepian_1978, Xu_1984}. Together with the majorization condition in our Theorem~\ref{theorem:rQFT_low}, this implies that there exists a number $k_0$ such that $\lambda_{n,j}^k \ge \tilde{\lambda}_{j}^k$ for $k < k_0$ and $\lambda_{n,j}^k \leq \tilde{\lambda}_{j}^k$ for $k \ge k_0$. With this, we will then prove that $\lambda_{n,j}^0 \ge \tilde{\lambda}_{j}^0$ and $\lambda_{n,j}^1 \le \tilde{\lambda}_{j}^1$, thus $k_0 = 1$ and any upper bounds on non-leading eigenvalues of $\tilde{T}_{j}$ can be directly applied to $T_{n,j}$.

To prove $\lambda_{n,j}^0 \ge \tilde{\lambda}_{j}^0$, we treat $n$ as a continuous variable and $T_{n,j}$ as smoothly dependent on $n$. By the Hellmann-Feynman theorem, we have
\begin{equation}
    \frac{d\lambda_{n,j}^k}{dn} = \left(v_{n,j}^{k}\right)^T \frac{dT_{n,j}}{dn} v_{n,j}^{k}
\end{equation}
where the derivatives of matrix elements of $T_{n,j}$ are
\begin{equation}
    \begin{split}
        &\frac{dT_{n,j}[x, y]}{dn}
        =\frac{d}{dn}\left( \frac{1}{2^n} \frac{\text{sin}(\pi(x-y)/2^j)}{\text{sin}(\pi(x-y)/2^n)} \right)\\
        = & \frac{\log (2)}{2^n} \frac{\sin \left(\pi (x-y)/2^j\right)}{\sin \left(\pi (x-y)/2^n\right)} \left(\frac{\pi(x-y)/2^n} {\tan \left(\pi(x-y)/2^n\right)} - 1\right)
    \end{split}
    \label{eq:dA_dn_elements}
\end{equation}
It is easy to see that $dT_{n,j}[x,y]/dn$ is strictly negative besides the diagonal elements being zero, i.e. $dT_{n,j}[x,x]/dn = 0$ and $dT_{n,j}[x,y]/dn < 0$ for $x \neq y$. We also observe that $T_{n,j}[x,y] > 0$ for all $x,y$, thus by the Perron-Frobenius theorem, the first eigenvector $v_{n, j}^{0}$ can be chosen to have strictly positive components. This means that
\begin{equation}
    \frac{d\lambda_{n,j}^0}{dn} = \left(v_{n, j}^{0}\right)^T \frac{dT_{n,j}}{dn} v_{n, j}^{0} \le 0
\end{equation}
Thus as $n$ increases, the first eigenvalue cannot increase, resulting in the inequality $\lambda_{n,j}^0 \ge \tilde{\lambda}_{j}^0$. 

Then we prove $\lambda_{n,j}^1 \leq \tilde{\lambda}_{j}^1$. To do this, we employ three facts. Firstly, We observe that $T_{n,j}$ is a Toeplitz matrix, i.e. $T_{n,j}[x,y] = T_{n,j}[x+1, y+1]$ for $x,y \in [0,2^j-2]$, and it is symmetric. This suggests all the eigenvectors are either symmetric or anti-symmetric, depending on the parity:
\begin{equation}
    v_{n,j}^{k}[x] = 
    \begin{cases}
        v_{n,j}^{k}[2^j-1-x] &\text{if} ~ k = 0, 2, 4...\\
        -v_{n,j}^{k}[2^j-1-x] &\text{if} ~ k = 1, 3, 5...\\
    \end{cases}
\end{equation}
where $v[x]$ denotes the $(x+1)$-th element of the vector $v$. Therefore, applying any symmetric Toeplitz matrix on $v_{n, j}^{1}$ must also produce an anti-symmetric vector. Secondly, it is known that $v_{n, j}^{k}$ has exactly $k$ nodes, i.e. $\text{sign}(v_{n, j}^{k}[x]) = -\text{sign}(v_{n, j}^{k}[x+1])$ occurs $k$ times among $x \in [0,2^{j}-2]$, where $\text{sign}(v[x])$ denotes the sign of the vector element $v[x]$ \cite{Xu_1984}. Therefore, we can choose $v_{n, j}^{1}$ to be non-negative in the first half and non-positive in the second half. Thirdly, we note that using Eq.~\eqref{eq:dA_dn_elements} one can verify that for the range $x, y \in [0,2^{j-1}-1]$,
\begin{equation}
     \frac{dT_{n,j}}{dn}[x, 2^j-1-y] \leq \frac{dT_{n,j}}{dn}[x, y] \leq 0
\end{equation}
Combing all three facts, we can conclude that the first half of $(dT_{n,j}/dn \cdot v_{n, j}^{1})$ is non-negative, i.e. for $x \in [0,2^{j-1}-1]$,
\begin{equation}
\begin{split}
     & \left(\frac{dT_{n,j}}{dn} \cdot v_{n, j}^{1} \right)[x]\\
    =& \sum_{y=0}^{2^j-1} \frac{dT_{n,j}}{dn}[x, y] v_{n, j}^{1}[y]\\
     =& \sum_{y=0}^{2^{j-1}-1} \left(\frac{dT_{n,j}}{dn}[x, y] - \frac{dT_{n,j}}{dn}[x, 2^j-1-y]\right) v_{n, j}^{1}[y]\\
     \ge& 0
\end{split}
\end{equation}
because $(dT_{n,j}/dn \cdot v_{n, j}^{1})$ is anti-symmetric, the second half $x \in [2^{j-1},2^j-1]$ of it will be non-positive. Thus for any $x \in [0,2^j-1]$,
\begin{equation}
    \text{sign}\left(v_{n, j}^{1}[x]\right) = \text{sign}\left(\left(\frac{dT_{n,j}}{dn} v_{n, j}^{1}\right)[x]\right)
\end{equation}
This means that
\begin{equation}
    \begin{split}
        \frac{d\lambda_{n, j}^1}{dn} &= \left(v_{n, j}^{1}\right)^T \frac{dT_{n,j}}{dn} v_{n, j}^{1} \\
    &= \sum_{x=0}^{2^{j}-1} \left(\frac{dT_{n,j}}{dn}v_{n, j}^{1} \right)[x] ~ v_{n, j}^{1}[x] \\
        &\ge 0
    \end{split}
\end{equation}
Thus as $n$ increases, the second eigenvalue cannot decrease. Therefore it must be the case that $\lambda_{n, j}^1 \leq  \tilde{\lambda}_{j}^1$.

\section{Entanglement measures of the QFT}
\label{appendix:opEE_EP}
In this appendix, we show that any entanglement measure of $Q_n$ and $Q_n^P$ are identical. 
Recall that $Q_n$ is the bit-reserved QFT operator and $Q_n^P$ is $Q_n$ without $H$ gates. To be more general, for any unitary operator $U_n$ acting on $n$ qubits satisfying the following decomposition for any $j$,
\begin{equation}
    U_n = (U^{1}_{j} \otimes U^{2}_{n-j}) \Omega_{n,j} (U^{3}_{j} \otimes U^{4}_{n-j})
    \label{eq:Un_decomp_app}
\end{equation}
where $U^{1}_{j}, U^{3}_{j}$ and $U^{2}_{n-j}, U^{4}_{n-j}$ are some unitary gates acting on the first $j$ qubits and the last $n-j$ qubits respectively, then any entanglement measure of $U_n$ will be the same to those of $Q_n$. The operator $Q_n^P$ belongs to $U_n$ because it can be expressed as
\begin{equation}
    Q_n^P = (Q_j^P \otimes I_{n-j}) \Omega_{n,j} (I_j \otimes Q_{n-j}^P)
\end{equation}
which is very similar to the recursive structure of $Q_n$ except that we set $Q_1^P = I$ instead of $H$.

We first elaborate on what we mean by an entanglement measure of an operator. Generally, there are two types, one being state-independent measures which consider the operators themselves as being entangled in their Hilbert space, such as the operator entanglement entropy \cite{Bandyopadhyay_2005}, and the other being state-dependent measures which consider how much entanglement the operator can generate on some quantum state, such as the entangling power \cite{Zanardi_2000}. 

Both types of measures are inspired by or based on entanglement measures of a quantum state, which has a long history of study. In the context of states, entanglement measures are formally defined as a non-negative real function that cannot increase under local operations and classical communication (LOCC), and should evaluate to zero on separable states \cite{Plenio_2006}. A property directly followed by the definition is that local unitaries do not change the entanglement measure. One of the most common entanglement measures is the von Neumann entropy: consider a quantum state $|\psi\rangle$ on some qubits and its Schmidt decomposition
\begin{equation}
    |\psi\rangle = \sum_k \alpha_k |a_k\rangle \otimes |b_k\rangle
\end{equation}
where the Hilbert space is partitioned into two subsystems (i.e. two sets of qubits), $\mathcal{A}$ and $\mathcal{B}$, with $\{|a_k\rangle\}$ and $\{|b_k\rangle\}$ being states on each subsystem respectively. The von Neumann entropy $S(|\psi\rangle)$ is then
\begin{equation}
    S(|\psi\rangle) = - \sum_k \alpha_k^2 \log(\alpha_k^2) 
\end{equation}
Inspired by this, we can combine the input space and the output space of an operator to transform it into an entangled state in a doubled Hilbert space. The state-independent entanglement measure of the operator is thus any usual entanglement measure on the resulting state. For example, consider the Schmidt decomposition of a unitary operator
\begin{equation}
    U = \sqrt{N} \sum_{k} \sigma_k A_k \otimes B_k
\end{equation}
where the normalized Schmidt coefficients $\{\sigma_k\}$ are identical to that of the resulting state. The von Neumann operator entanglement entropy can be then defined as
\begin{equation}
    K_S(U) = - \sum_k \sigma_k^2 \log(\sigma_k^2) 
\end{equation}
These types of measurement are most valuable for examining the expressibility of an operator as a matrix product operator. To see why all state-independent entanglement measures of $Q_n$ and $U_n$ are identical, recall the decomposition
\begin{equation}
    Q_n = \left(I_j \otimes Q_{n-j}\right) \Omega_{n,j} \left(Q_{j} \otimes I_{n-j}\right)
    \label{eq:QFT_decomp_app}
\end{equation}
together with Eq.~\eqref{eq:Un_decomp_app} one can see $U_n$ and $Q_n$ are both $\Omega_{n,j}$ transformed under local unitaries, thus the LOCC principle ensures all state-independent entanglement measures on $Q_n$ and $U_n$ are the same.

The other type of entanglement measure of an operator considers how much an entanglement measure of a state can be changed by the operator. For example, for some entanglement measure on a state $E(|\psi\rangle)$, we can measure its maximal change under the transformation $U$ as
\begin{equation}
    K_{\max}(U) = \sup_{|\phi\rangle} | E(U|\phi\rangle) - E(|\phi\rangle) |
\end{equation}
or the average change as
\begin{equation}
    K_{\text{avg}}(U) = \overline{| E(U|\phi\rangle) - E(|\phi\rangle)|}^\phi
\end{equation}
Many works in the literature have examined the two quantities under different definitions, such as which $E$ to use, whether $|\phi\rangle$ starts as product states, or whether there are ancillary systems in $|\phi\rangle$ so its Hilbert space is larger than that of $U$ acts on, etc. \cite{Dur_2001, Bennett_2003, Nielsen_2003, Wang_2003, Rajarshi_2018}. Under all variants of the definitions, these types of measures are identical between $Q_n$ and $U_n$. This is because, for any state $|\phi_1\rangle$, one can always bijectively map it to another state $|\phi_2\rangle$ in the same Hilbert space, so that for any entanglement measures $E$,
\begin{align}
    &E(|\phi_1\rangle) = E(|\phi_2\rangle) \label{eq:E_map_1} \\
    &E(Q_n|\phi_1\rangle) = E(U_n|\phi_2\rangle)
    \label{eq:E_map_2}
\end{align}
Therefore, the two operators essentially form the same set of entanglement entropy change among all states $|\phi\rangle$ in any Hilbert space $\mathcal{H}$, i.e.
\begin{equation}
\begin{split}
    &\{|E(Q_n|\phi\rangle) - E(|\phi\rangle)| ~{\big |}~ |\phi\rangle \in \mathcal{H} \} \\ = &\{|E(U_n|\phi\rangle) - E(|\phi\rangle)| ~{\big |}~ |\phi\rangle \in \mathcal{H} \}
\end{split}
\end{equation}
and their supremum, maximum, or average are identical. To find the map between $|\phi_1\rangle$ and $|\phi_2\rangle$, one can see from Eq.~\eqref{eq:QFT_decomp_app} that $Q_{n-j}$ is a local unitary, therefore the entanglement measure $E(Q_n |\phi_1\rangle)$ can be reduced to
\begin{equation}
    E(Q_n |\phi_1\rangle) = E\left(\Omega_{n, j} (Q_{j} \otimes I_{n-j})|\phi_1\rangle\right)
\end{equation}
Then using Eq.~\eqref{eq:Un_decomp_app} one can obtain the exactly same entanglement measure by applying ${U_j^3}^\dagger, {U_{n-j}^4}^\dagger$ and $Q_{j}$ on $|\phi_1\rangle$ before applying $U_n$,
\begin{equation}
\begin{split}
    &E\left(U_n \left({U_j^3}^\dagger \otimes {U_{n-j}^4}^\dagger \right) \left(Q_{j} \otimes I_{n-j} \right)  |\phi_1\rangle \right)\\ 
    = &E\left( \left({U_j^1} \otimes {U_{n-j}^2} \right) \Omega_{n,j} \left(Q_{j} \otimes I_{n-j} \right)  |\phi_1\rangle \right)\\ 
    = &E\left( \Omega_{n,j} \left(Q_{j} \otimes I_{n-j} \right)  |\phi_1\rangle \right)\\ 
\end{split}
\end{equation}
and thus we have
\begin{equation}
    |\phi_2\rangle = \left({U_j^3}^\dagger \otimes {U_{n-j}^4}^\dagger \right) \left(Q_{j} \otimes I_{n-j} \right) |\phi_1\rangle
\end{equation}
which satisfies Eq.~\eqref{eq:E_map_2}. Eq.~\eqref{eq:E_map_1} is also satisfied because Here ${U_j^3}^\dagger, {U_{n-j}^4}^\dagger$ and $Q_{j}$ are all local unitaries. Notice this map is $j$-dependent, thus for different $j$ one needs to use a different set of unitaries. This concludes that all state-dependent types of entanglement measures on $Q_n$ and $U_n$ are equivalent.

This connection allows us to apply any entanglement measure of $U_n$ (specifically $Q_n^P$) directly to $Q_n$. As demonstrated in Section~\ref{subsection:area_law}, $Q_{n}^P$ can be expressed as
\begin{equation}
    Q_{n}^{P} = \exp(i\pi\sum_{l=1}^{n-1}\sum_{m=l+1}^n \left(\frac{1}{2}\right)^{m-l} |1 \rangle\langle 1|_{l} \otimes |1 \rangle\langle 1|_{m})
\end{equation}
which can also be interpreted as the time evolution under Hamiltonian $H_n$ with $\pi$ time unit:
\begin{equation}
    H_n = -\sum_{l=1}^{n-1}\sum_{m=l+1}^n \left(\frac{1}{2}\right)^{m-l} |1 \rangle\langle 1|_{l} \otimes |1 \rangle\langle 1|_{m}
\end{equation}
In \cite{Gong_2017}, the author proved the following: for a quantum state $|\psi\rangle$ on a $D$-dimensional finite or infinite lattice, if it time-evolves under any Hamiltonian with two-site interactions satisfying
\begin{equation}
    H_n = \sum_{l,m} h_{l,m}, \quad \norm{h_{l,m}} \leq \frac{1}{r^{D+1}}
\end{equation}
where $r$ is the distance between two sites, $h_{l,m}$ is a Hermitian operator acting on two sites, and $\norm{h_{l,m}}$ denotes its operator norm (i.e. the largest magnitude of an eigenvalue), the rate of the change of von Neumann entropy with respect to some subregion $\mathcal{V}$, known as the entanglement rate, is at most proportional to the size of the boundary of $\mathcal{V}$, which we denote as $\partial \mathcal{V}$. That is,
\begin{equation}
    \left| \frac{dS_{\mathcal{V}}(|\psi(t)\rangle)}{dt} \right| \leq O(\partial \mathcal{V})
    \label{eq:dS_dt}
\end{equation}
For the case of $Q_{n}^P$ on a 1D chain of qubits, each term has operator norm
\begin{equation}
    \left\| \left(\frac{1}{2}\right)^{m-l} |1 \rangle\langle 1|_{l} \otimes |1 \rangle\langle 1|_{m} \right\| = \frac{1}{2^{r}}
\end{equation}
which is exponentially smaller than $1/r^2$, thus the entanglement rate is bounded by a constant, i.e.
\begin{equation}
    \left| \frac{dS_{j}(|\psi(t)\rangle)}{dt} \right| \leq O(1)
\end{equation}
where $S_j$ denotes von Neumann entropy calculated at the partition between qubit $j$ and $j+1$. Therefore, the change of entanglement by evolving $\pi$ time unit is also bounded by a constant, i.e.
\begin{equation}
\begin{split}
    \Delta S_j 
    = & \left| S_j(Q_n^{P}|\psi\rangle) - S_j(|\psi\rangle) \right| \\
    = &\int_{0}^{\pi} \frac{dS_{j}(|\psi(t)\rangle)}{dt} dt \\
    \leq &O(1)
\end{split}
\end{equation}
which serves as an upper bound on the entanglement $Q_n^P$ and $Q_n$ can generate on any input states. While this is not proof of the $Q_n$'s exponentially decaying Schmidt coefficients, it gives some intuition as to why $Q_n$ can only generate small entanglement.

\section{Schmidt Coefficients from QFT-TN}
\label{appendix:schmidt_from_QFT_TN}
In this appendix, we obtain the QFT's Schmidt coefficients through an alternative approach: the rewriting of the QFT tensor network. We present a special orthogonality property that is intrinsic to the QFT-TN, which can be used to significantly simplify the diagram for extracting Schmidt coefficients and obtain the same results with the main theorem.

We first introduce the more general family of copy tensors and phase tensors. These tensors do not appear in QFT-TN but they are useful for rewriting the tensor diagram. A general copy tensor can have an arbitrary number of indices, for example,
\begin{equation}
   \input{copy_general.tikz}
   \label{eq:copy_general}
\end{equation}
and its tensor element is 1 when all indices have the same value and 0 otherwise. For phase tensors, the number of indices is limited to $\{0, 1, 2, 3, 4\}$. In Section~\ref{subsection:QFT_TN}, we have already introduced the order-3 and order-4 phase tensors
\begin{equation}
    \input{3_4-phase.tikz}
\end{equation}
The order-2 phase tensors have two types:
\begin{equation}
    \input{2-phase.tikz}
\end{equation}
where the second type is simply the phase gate $P(\theta)$. The order-1 phase tensor is a vector defined as
\begin{equation}
    \input{1-phase.tikz}
\end{equation}
and an order-0 phase tensor is the constant $e^{i\theta}$.
All the phase tensors and copy tensors preserve the definition under rotation and reflection.

\begin{figure}[tb]
    \centering
    \includegraphics[width=1\columnwidth]{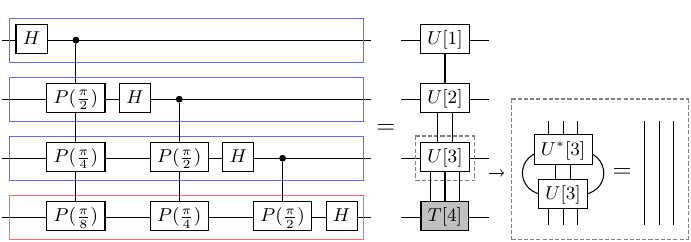}
    \caption{The orthogonal property of a QFT-TN. By treating each sequence of tensors acting on the same qubit as a single tensor, the QFT-TN corresponds to a right-canonical MPO with bond dimension un-optimized (here ``right'' means ``down'' in the orientation of the quantum circuit). In this 4-qubit example, the orthogonal property of $U_3$ is shown, and $U_1$, $U_2$ obey the same identity, but $T_4$ does not and thus it is the orthogonality center.}
    \label{fig:qft_right_orthogonal}
\end{figure}

Now we show several identities associated with copy tensors and phase tensors. Two copy tensors can be combined into one:
\begin{equation}
    \input{copy_identity.tikz}
    \label{eq:copy_identity}
\end{equation}
By induction, the contracted product of an arbitrary number of general copy tensors with $n$ total open indices can be merged into a single copy tensor with $n$ open indices. A copy tensor with 2 open indexes is identity. Phase tensors have properties that allow themselves to cancel each other (ie. contracted to identity). Recall from Section~\ref{subsection:QFT_TN} that an order-4 tensor is an isometric tensor. This means an order-4 phase tensor contracted to another order-4 phase tensor with negated parameters (conjugate of itself) by one of the horizontal indices and one of the vertical indices gives the identity tensor
\begin{equation}
    \input{4-phase_identity.tikz}
    \label{eq:4-phase_identity}
\end{equation}
Similarly, two order-3 phase tensors give identity up to a constant of 2,
\begin{equation}
    \input{3-phase_identity.tikz}
    \label{eq:3-phase_identity}
\end{equation}
The order-2 phase tensors representing phase gates can be combined,
\begin{equation}
    \input{2-phase_identity.tikz}
    \label{eq:2-phase_identity}
\end{equation}
and if $\theta_1 = -\theta_2$ they cancel to identity. We also note that the $H$ gate is both symmetric and unitary, which can also be represented as a tensor diagram
\begin{equation}
    \input{h-tensor_identity.tikz}
    \label{eq:h-tensor_identity}
\end{equation}
By combing Eq.~\eqref{eq:copy_identity}, Eq.~\eqref{eq:4-phase_identity}, and Eq.~\eqref{eq:h-tensor_identity}, we can further construct identities associated with the entire sequence of tensors acting on the a qubit,
\begin{equation}
    \resizebox{0.88\hsize}{!}{\input{layer_isometry.tikz}}
    \label{eq:layer_isometry}
\end{equation}
where we can see the sequence of tensors on the top with negative parameters are the conjugate transpose of the sequence of tensors on the bottom, thus the bottom sequence of tensors forms an isometric matrix by bending qubit indices upwards. This gives the QFT-TN an orthogonal property similar to the right-canonical form of MPO, which is demonstrated in Fig.~\ref{fig:qft_right_orthogonal}. That is, if we consider the sequence of tensors on the same qubit as a single tensor and view them as a site on an MPO, the MPO will have the orthogonality center at the last qubit. We note this orthogonal property is the same as treating it as a right-canonical MPS with doubled Hilbert space, by bending the input and output indices into the same direction.

\begin{figure}[tb]
    \centering
    \includegraphics[width=1\columnwidth]{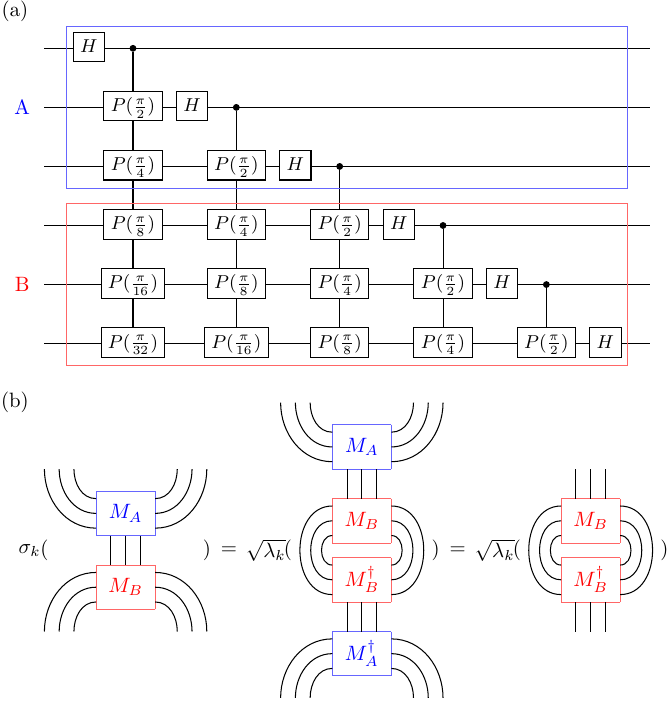}
    \caption{(a) An example of partitioning 6-qubit QFT-TN at the center into two sets of tensors $A$ and $B$. (b) Diagrammatic illustration of simplifying Schmidt coefficients calculations. $M_A$ and $M_B$ represent tensors in $A$ and $B$ respectively. Since all tensors in $M_A$ are isometric, singular values can be calculated only from $M_B$}
    \label{fig:qft_partition}
\end{figure}

We also introduce a useful method to reason about phase tensors, that is the input value of their indices can be inserted into their parameters, and the indices can be slid across the tensor. For example, the order-4 phase tensors can be re-expressed as
\begin{equation}
    \input{4-phase_insert.tikz}
    \label{eq:4-phase_insert}
\end{equation}
and the order-3 phase tensors can be re-expressed as
\begin{equation}
    \input{3-phase_insert.tikz}
    \label{eq:3-phase_insert}
\end{equation}
To see an example of how these are useful, Eq.~\eqref{eq:4-phase_identity} and Eq.~\eqref{eq:3-phase_identity} can be derived with them without writing out tensor elements.
\begin{figure*}[tb]
    \centering
    \includegraphics[width=2\columnwidth]{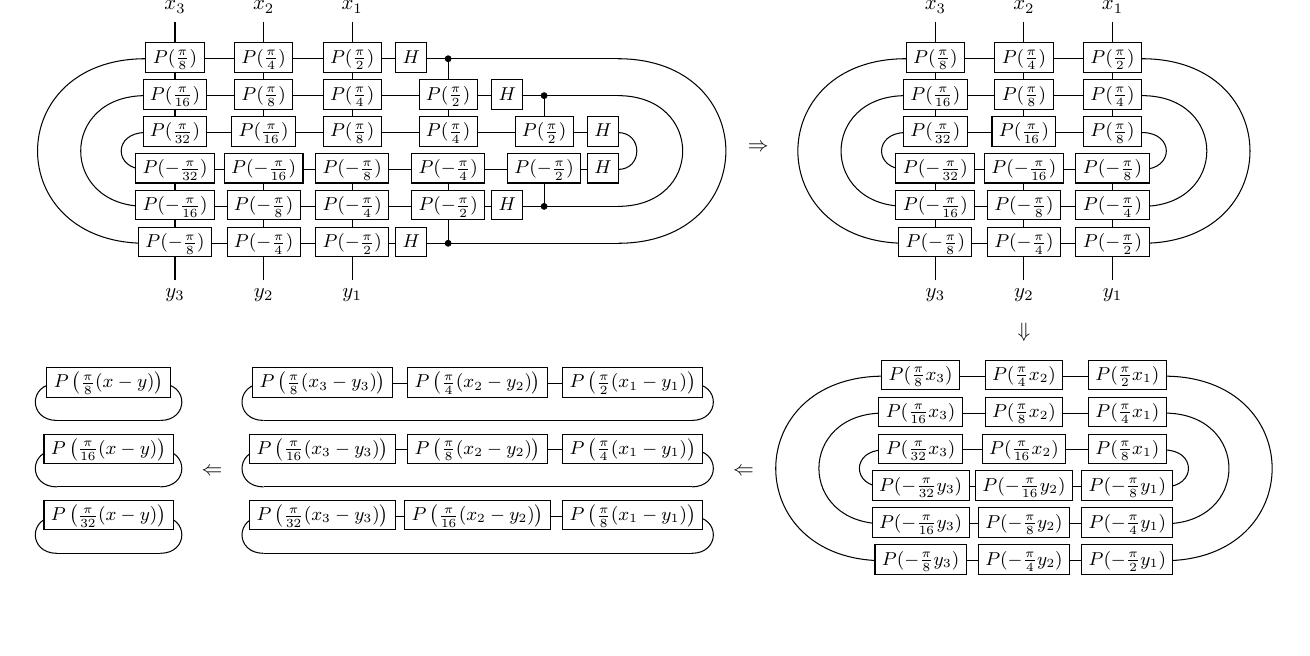}
    \caption{Diagrammatic illustration of simplifying matrix elements of $M_B M_B^\dagger$, with $M_B$ shown in Fig.~\ref{fig:qft_partition}. The first diagram to the second is due to all $H$ gates and phase gates being unitaries. The second to the third employs Eq.~\eqref{eq:4-phase_insert} and Eq.~\eqref{eq:3-phase_insert}. The third to the fourth employs Eq.~\eqref{eq:2-phase_identity}. The fourth to the last uses Eq.~\eqref{eq:2-phase_identity} again and writes $x = 2^2 x_1+ 2 x_2 + x_3$ and $y = 2^2 y_1+ 2 y_2 + y_3$.}
    \label{fig:qft_reduction}
\end{figure*}

Now we use these properties of the tensor pieces to extract the QFT's Schmidt coefficients. Consider the partition of $n$-qubit QFT into two systems $\mathcal{A}$ with qubits 1 to $j$ and $\mathcal{B}$ with qubits $j+1$ to $n$. We bend qubit indexes (input and output) in $\mathcal{A}$ onto one direction and qubit indexes in $\mathcal{B}$ onto the other direction, resulting in a matrix $M$ with dimension $(2^j)^2 \times (2^{n-j})^2$. The operator Schmidt decomposition is thus the same as the singular value decomposition of $M$. Additionally, $M$ is the product of two matrices $M_A M_B$ corresponding to the tensors on each partition. From Fig.~\ref{fig:qft_right_orthogonal} we can show $M_A$ is isometric ($M_A^\dagger M_A = I$) and the orthogonality center is contained in $M_B$, thus the singular values $\sigma(M)$ is the same as the square root of eigenvalues $\sqrt{\lambda}(MM^\dagger) = \sqrt{\lambda}(M_BM_B^\dagger)$. To illustrate diagrammatically, a 6-qubit example with a partition at the center is shown in Fig.~\ref{fig:qft_partition}. The matrix elements of $M_BM_B^\dagger$ can be easily calculated: firstly we observe all the $H$ gates and phase gates in $M_B$ will cancel those in $M_B^\dagger$, then we can use the identities of phase tensors to simplify the diagram significantly as shown in Fig.~\ref{fig:qft_reduction}. The last step shows that the matrix element $(M_BM_B^\dagger)[x,y]$ is the product $\prod_{k=1,2,3}\Tr\left(P\left(\pi(x-y)2^k/2^6\right)\right)$, where $P(\theta)$ is the phase gate defined in Fig.~\ref{fig:qft_circ}. It is easy to generalize that for $n$-qubit QFT with a partition between qubit $j$ and $j+1$:
\begin{equation}
    (M_B M_B^\dagger)[x,y] = \prod_{k=1,2,...n-j}\Tr\left(P\left(\pi(x-y)2^k/2^n\right)\right)
\end{equation}
which can be turned into a sum of geometric series and simplified as
\begin{equation}
    \begin{split}
       (M_B M_B^\dagger)[x,y]
        = & \prod_{k=1}^{n-j} \left(1 + e^{i\pi(x-y)2^k/2^n}\right)\\ 
        = & \sum_{k=0}^{2^{n-j}-1} e^{i\pi(x-y)k/2^{n-1}}\\
        = & \frac{1-e^{i\pi(x-y)2^{n-j}/2^{n-1}}}{1-e^{i\pi(x-y)/2^{n-1}}}\\
        = & e^{i\pi(x-y)(2^{n-j}-1)/2^n}\frac{\text{sin}(\pi(x-y)/2^j)}{\text{sin}(\pi(x-y)/2^n)}
    \end{split}
    \label{eq:geo_gen}
\end{equation}
which is identical to Eq.~\eqref{eq:A_elements_simplified} in Section~\ref{subsection:QFT_LE_proof}. The Schmidt coefficients can thus be calculated by solving the eigenvalues of this matrix.

\begin{figure}[tb]
    \centering
    \includegraphics[width=1\columnwidth]{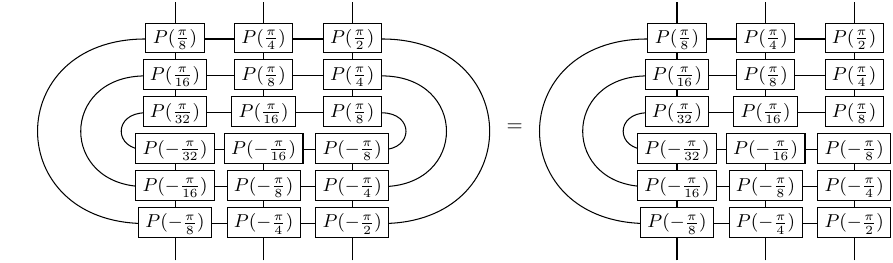}
    \caption{An example of simpliying $M_B M_B^\dagger$ to $2^n F_{n,j} F_{n,j}^\dagger$ with $n=6$ and $j=3$. The diagram on the left comes from step two in Fig.~\ref{fig:qft_reduction}.}
    \label{fig:qft_submatrix_reduction}
\end{figure}

One may wonder how is this connected to our previous analysis in Section~\ref{subsection:QFT_LE_proof} that the Schimdt coefficients of $Q_n$ at cut $j$ are the same as singular values of $F_{n,j}$ (the top $2^j \times 2^{n-j}$ submatrix of $F_n$). We will demonstrate this by showing that $M_B M_B^\dagger = 2^n F_{n,j}F_{n,j}^\dagger$. We first claim that the following tensor diagram represents the $2^m \times 2^l$ top left submatrix of $\sqrt{2^n} F_n$:
\begin{equation}
    \input{QFT_submatrix.tikz}
    \label{eq:QFT_submatrix}
\end{equation}
That is, the tensor network evaluates to $\exp(i2\pi x y/2^n)$ where $x = 2^{l-1} x_1 + ... + x_l$ and $y = 2^{m-1} y_1 + ... + y_m$. We will refer to this tensor network as the generalized QFT-TN. It is easy to see that the tensor network of $M_B M_B^\dagger$ can be simplified to this tensor network contracted with its conjugate transpose with $m=j$ and $l=n-j$, or equivalently $M_B M_B^\dagger = 2^n F_{n,j}F_{n,j}^\dagger$; see Fig.~\ref{fig:qft_submatrix_reduction} for an example. One can achieve this simplification by iteratively applying the identity to all order-4 and order-3 phase tensors, i.e.
\begin{equation}
    \input{2_3-phase_decompose_to_copy.tikz}
\end{equation}
Therefore, for any $n$ and $j$, the lower partition $M_B$ has the same set of non-zero singular values with those of $\sqrt{2^n} F_{n,j}$, linking back to results in Section~\ref{subsection:QFT_LE_proof}.

\section{Emergent QFT Quantum Circuit}
\label{appendix:emergent_QFT_circuit}
In this appendix, we show that the quantum circuits of the QFT can be derived from the tensor network structure introduced in Eq.~\eqref{eq:QFT_submatrix}, which is a generalized QFT-TN that represents any submatrix of the QFT. To demonstrate, consider the 4-qubit example, where $l=m=4$. Eq.~\eqref{eq:QFT_submatrix} becomes
\begin{equation}
    \input{QFT_submatrix_4q.tikz}
    \label{eq:QFT_submatrix_4q}
\end{equation}
We introduce a few identities to remove phase gates with parameters $2k\pi$, with $k$ being an integer. Firstly, an order-3 phase gate will reduce to identity with an extra plus tensor,
\begin{equation}
    \input{3-phase_2kpi.tikz}
\end{equation}
where the plus tensor is defined as
\begin{equation}
    \input{plus.tikz}
\end{equation}
The order-4 phase tensor will reduce to two crossing identity tensors
\begin{equation}
    \input{4-phase_2kpi.tikz}
\end{equation}
and any order-4 phase tensor can eat up a plus tensor to become an order-3 tensor
\begin{equation}
    \input{4-phase_plus.tikz}
\end{equation}
This allows us to reduce Eq.~\eqref{eq:QFT_submatrix_4q} to
\begin{equation}
    \input{QFT_binary_decomp_simplify_lines.tikz}
    \label{eq:QFT_binary_decomp_simplify_lines}
\end{equation}
It is also easy to see that the order-3 phase tensors have the identity
\begin{equation}
    \input{2-phase_pi_identity.tikz}
\end{equation}
and thus $H$ gates and copy tensors naturally arise from such decomposition. Ignoring the scalar, we then have the diagram 
\begin{equation}
    \input{QFT_circuit_bend.tikz}
\end{equation}
and by bending $y$ index lines to the horizontal direction, we obtain the QFT quantum circuit
\begin{equation}
    \input{QFT_circuit_tn.tikz}
\end{equation}

We also consider how would one get the QFT circuit if Eq.~\eqref{eq:QFT_submatrix} was not known. The first step one may do is to define a generalized version of the order-2 phase tensors
\begin{equation}
    \input{general_2-phase.tikz}
\end{equation}
with both indices $x$ and $y$ of arbitrary dimension (One may not call it a phase tensor, and may not write indices in two directions, but we do so here for consistency with our previous definitions). It is easy to see that the Fourier matrix $F_n$ scaled by $2^{n/2}$ is equivalent to one of such tensor with $\theta = \pi/2^{n-1}$ and $x, y \in [0, 2^{n-1}]$, i.e.
\begin{equation}
    \input{QFT_as_2-phase.tikz}
    \label{eq:QFT_as_2-phase}
\end{equation}
Again, we use $n=4$ as an example to show a derivation of the QFT's quantum circuit through the decomposition of the generalized order-2 phase tensor. For convenience, we consider two indices $x$ and $y$ as variables in the phase parameter, i.e. applying the insertion identity
\begin{equation}
    \input{QFT_as_2-phase_4q.tikz}
\end{equation}
Then one can express $x$ and $y$ in terms of their binary expansions
\begin{equation}
    \resizebox{1\hsize}{!}{\begin{tikzpicture}
	\begin{pgfonlayer}{nodelayer}
		\node [style=box] (2) at (0, 0) {$P(\frac{\pi}{2^3}(2^3x_1 + 2^2x_2 + 2^1x_3 + 2^0x_4)(2^3y_1 + 2^2y_2 + 2^1y_3 + 2^0y_4))$};
	\end{pgfonlayer}
\end{tikzpicture}
}
\end{equation}
Through our definition of the phase tensor, we can then break the sum into a product of phase tensors
\begin{equation}
    \resizebox{1\hsize}{!}{\input{QFT_binary_decomp.tikz}}
\end{equation}
which is essentially breaking the exponential of the sum of phase parameters into a product of the exponential of each phase parameter. Then one observes that by bringing the index lines back, one can get Eq.~\eqref{eq:QFT_submatrix_4q}. Equivalently, one observes phase tensors with parameters being multiple of $2\pi$ evaluate to constant $1$, thus they can be removed from the diagram and we can directly achieve Eq.~\eqref{eq:QFT_binary_decomp_simplify_lines}.

\section{Stability of the Zip-up Algorithm over the QFT-TN}
\label{appendix:zip_up_stable}
In this appendix, we make some arguments that the zip-up algorithm is completely stabilized for the QFT-TN. While we do not have proof, the tensor diagram analysis and numerical simulation strongly suggest so, thus we conjecture it is true for the QFT-TN with any number of qubits.

We first explain what we mean by being stabilized. For the contraction of two MPOs, The zip-up algorithm does local SVD at each step, where the singular values do not reflect the actual Schmidt coefficients of the contracted MPO. Therefore, if one does truncation over the singular values, one may generally not achieve optimal truncation, and an error larger than the cutoff would be introduced. However, if one can form isometric tensors by combing each site of the two MPOs before the truncation, the isometries will allow local SVDs to reflect the whole system's Schmidt coefficients, therefore truncating local singular values are optimal. We call this being completely stabilized. For contracting two generic MPOs, one may not expect this property, thus the original algorithm in \cite{Stoudenmire_2010} heuristically aligns the orthogonality centers to somewhat stabilize the algorithm. 

We then explain why the QFT-TN is likely to get completely stabilized. Consider the situation where one tries to contract the $j_0$-th intermediate MPO and the next phase MPO with an $H$ gate sandwiched in between, which can be demonstrated by the diagram
\begin{equation}
    \input{interMPO_with_next_phase.tikz}
\end{equation}
where $\{U[k], T[n]\}$ are sites of the $j_0$-th intermediate MPO. As introduced before, to completely stabilize the zip-up algorithm, one wants to show that combining two sites besides the orthogonality centers results in the isometric tensors
\begin{equation}
    \input{interMPO_with_next_phase_ortho.tikz}
\end{equation}
This is already true for sites $j_0+2$ to $n-1$, i.e. sites with phase tensors, since one can apply Eq.~\eqref{eq:4-phase_identity} from Appendix~\ref{appendix:schmidt_from_QFT_TN} to reduce the following tensor diagram to identity
\begin{equation}
    \input{diagonal_site_next_phase.tikz}
\end{equation}
and for sites $1$ to $j_0$ this is trivial since they are untouched and one already has isometric tensors from previous iterations. Therefore, the only problematic site is $j_0+1$. That is, we cannot guarantee the following diagram is equivalent to two identity tensors
\begin{equation}
    \input{site_j0+1.tikz}
    \label{eq:site_j0+1}
\end{equation}
However, we will show that, if the tensor $U[j_0+1]$ is \emph{horizontally diagonal}, then the site $j_0+1$ with an $H$ gate and a copy tensor also forms an isometric tensor. By horizontally diagonal, it means by setting vertical indices of the tensor $U[k]$ to some values, the resulting matrix in the horizontal direction is always diagonal. Equivalently, one can describe it as the diagram
\begin{equation}
    \input{diagonal_site.tikz}
\end{equation}
where the $x$ tensor is a projector onto $|x\rangle$ with $x\in\{0,1\}$. This property allows us to express the contraction of $U[k]$ and its conjugate transpose as a sum
\begin{equation}
    \input{diagonal_site_contraction.tikz}
\end{equation}
which also equals identity since $U[k]$ is guaranteed to be isometric. One can then evaluate a diagram like Eq.\eqref{eq:site_j0+1} as a sum over two diagrams:
\begin{equation}
    \input{diagonal_site_next_copy.tikz}
    \label{eq:diagonal_site_next_copy}
\end{equation}
where whether $x$ equals 0 or 1, the right part of the diagram evaluates to identity up to a factor $1/2$ by introducing an extra sum
\begin{equation}
    \input{copy_to_identity.tikz}
\end{equation}
which is due to the fact that the square of any element of $H$ equals $1/2$. Therefore, one can simplify Eq.\eqref{eq:diagonal_site_next_copy} to
\begin{equation}
    \input{diagonal_site_next_copy_simplify.tikz}
\end{equation}
As we discussed, this entirely relies on $U[j_0+1]$ being horizontally diagonal. We conjecture this is true, because for the first $j_0$ phase MPOs the bottom part $k > j_0$ are all formed by horizontally-diagonal phase tensors. In addition, we have done numerical simulations for $n$ up to 64 qubits and verified this is the case. Since the QFT has site tensors approaching a fixed form as $n$ goes large, we expect this behavior to hold for larger $n$. Even if this does not hold, the sites $k \ne j_0+1$ will all form isometric tensors, thus the zip-up algorithm will not be very unstable in any case.

\section{Schmidt coefficients of intermediate MPOs}
\label{appendix:intermediate_MPOs}
Assume we have exactly contracted the first $j_0$ MPOs. Using the same techniques introduced in Section~\ref{subsection:QFT_LE_proof}, the Schmidt coefficients of the current MPO can be obtained by reducing to the operator
\begin{multline}
    \Omega_{n,j,j_0} = \sum_{x,x'} \exp (i\sum_{l=1}^{\min(j,j_0)} \sum_{m=1}^{n-j} x_{l} x'_{m} \frac{\pi}{2^{l+m-1}}) \times \\ |x_j...x_1\rangle \langle x_j...x_1| \otimes |x_1'...x_{n-j}'\rangle \langle x_1'...x_{n-j}'|
\end{multline}
and the eigenvalue value equation is
\begin{equation}
    \frac{1}{2^n}\sum_{y=0}^{2^{\min(j,j_0)}-1} \frac{\text{sin}(\pi(x-y)/2^j)}{\text{sin}(\pi(x-y)/2^n)} v_{n,j,j_0}^{k,y} = \lambda_{n,j,j_0}^{k} v_{n,j,j_0}^{k,x}
\end{equation}
By setting $2N'W = 2^{\min(j,j_0)-j}$ for results in Appendix~\ref{appendix:DPSS_upperbounds}, we conclude that for $\max(2, {\frac{e\pi}{4}}2^{\min(j,j_0)-j}) \leq k\leq 2^{\min(j,j_0)}-1$,
\begin{equation}
    \sqrt{\lambda_{n,j,j_0}^{k}} \leq \frac{1}{\sqrt{k}}\exp \left(-\frac{2k+1}{2}\log\left(\frac{4k+4}{2^{\min(j,j_0)-j} e\pi}\right)\right)
\end{equation}
For $j > j_0$, the Schmidt coefficients decay even faster than those of $Q_n$. For $j \le j_0$, the Schmidt coefficients are identical to those of $Q_n$ (which makes sense since the first $j_0$ qubits are never touched after the first $j_0$ MPOs are contracted).

\section{Error bounds on truncating the QFT-MPO}
\label{appendix:error_bounds}
In this appendix, we prove that both the average and worst-case error is bounded by $O(n e^{-\chi \log(\chi/3)}/\sqrt{\chi})$, even though there is an extra $\sqrt{2^n}$ factor in the operator Schmidt decomposition.

We will focus on the truncation of a single bond, since all the bonds should have the same order of error due to Theorem~\ref{theorem:rQFT_low}. The sum over all bonds upper bounds the error of the entire MPO, if one views it as double-legged MPS \cite{Schuch_2008, Oseledets_2011}. For a single bond at cut $j$, we define operator $D_{n,j}^{\chi}$ to be the Schmidt decomposition components after the $\chi$-th largest Schmidt coefficient:
\begin{equation}
    D_{n,j}^{\chi} = \sqrt{2^n} \sum_{k = \chi}^{\min(2^j,2^{n-j})-1} \sigma_{n,j}^{k} A_{n,j}^{k} \otimes B_{n,j}^{k}
\end{equation}
The single-bond average error, denoted as $\tilde{e}_{\text{avg}}$, is then defined as
\begin{equation}
     \tilde{e}_{\text{avg}} = \mathop{\mathbb{E}}_{\psi \sim \mu_H} \left[ \langle \psi| {D_{n,j}^{\chi}}^\dagger D_{n,j}^{\chi} |\psi\rangle \right] 
\end{equation}
where $\mu_H$ denotes the ensemble of Haar random states. Equivalently, this is related to the Forbenius norm of $D_{n,j}^{\chi}$ by 
\begin{equation}
    \tilde{e}_\text{avg} = \frac{1}{2^n}\|D_{n,j}^{\chi}\|_F^2 = \sum_{k \ge \chi} \left(\sigma_{n,j}^{k}\right)^2
\end{equation}
and is thus bounded by $O(e^{-\chi \log(\chi/3)}/\sqrt{\chi})$ due to Theorem ~\ref{theorem:rQFT_low}. The total average error $e_{\text{avg}}$ is then $O(n e^{-\chi \log(\chi/3)}/\sqrt{\chi})$. 
We note that for the average error, there are many other measures than $e_{\text{avg}}$ in the MPO framework. For example 
the normalized Hilbert-Schmidt inner product considered in \cite{Woolfe_2014}:
\begin{equation}
    \langle Q, Q' \rangle = \frac{1}{2^n} \Tr \left( Q^\dagger Q' \right)
\end{equation}
which is of the same order as $e_{\text{avg}}$, and the relative distance $\mathcal{D}(Q_n, Q_n')$ used in \cite{Guo_2022}:
\begin{equation}
    \mathcal{D}(Q_n, Q_n') = \frac{\|Q_n - Q_n' \|_F^2}{\|Q_n \|_F \|Q_n' \|_F}
\end{equation}
which relates to $e_{\text{avg}}$ by 
\begin{equation}
    \mathcal{D}(Q_n, Q_n') = \frac{\|Q_n \|_F }{\|Q_n' \|_F}~ e_{\text{avg}}
\end{equation}
and thus at worst scale $O(c(\chi)^n n e^{-\chi \log(\chi/3)}/\sqrt{\chi})$ where $c(\chi)$ decreases as $\chi$ increases and is between 1 and 2. This implies that at worst linear-scaling $\chi$ is needed to keep $\mathcal{D}(Q_n, Q_n')$ constant. All of these quantities are easy to benchmark numerically with MPOs, and $e_{\text{avg}}$ is particularly simple and representative since it tightly connects to the fidelity of the double-legged MPS arising from reshaping the MPO.

\begin{figure}[tb]
    \centering
    \includegraphics[width=1\columnwidth]{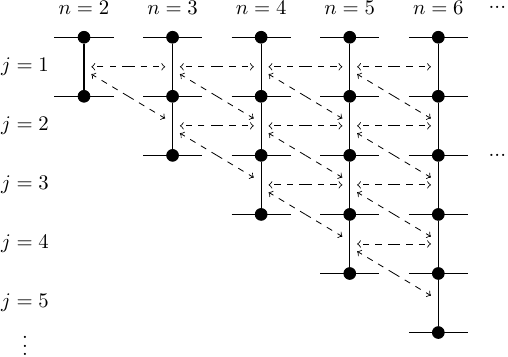}
    \caption{Diagrammatic illustration on how $e_{\max}(D_{n,j}^{\chi})$ are related for different $n$ and $j$. The horizontal lines denote that the error on bond $j$ has an upper bound independent of $n$, which is evident from Eq.~\eqref{eq:n-independent_emax}. The diagonal lines arise from the symmetry of the QFT, i.e. the Schmidt coefficients are identical at cut $j$ and $n-j$. Therefore, one can form a path between two bonds at any $n$ and $j$, thus the upper bound on $e_{\max}(D_{n,j}^{\chi})$ is also $j$-independent.}
    \label{fig:QFT_error_relation}
\end{figure}

We now consider the worst-case error for a single bond, which is defined as
\begin{equation}
     \tilde{e}_{\text{max}} = \sup_{|\psi\rangle:\langle \psi | \psi \rangle = 1} \langle \psi| {D_{n,j}^{\chi}}^\dagger D_{n,j}^{\chi} |\psi\rangle
\end{equation}
We first claim that $ \tilde{e}_{\text{max}}$ is the same as truncating $\Omega_{n,j}$, because they are transformed to each other by local unitary:
\begin{equation}
    Q_n = \left(I_j \otimes Q_{n-j}\right) \Omega_{n,j} \left(Q_{j} \otimes I_{n-j}\right)
\end{equation}
Therefore one can use the Schmidt vectors of $\Omega_{n,j}$ to upper bounds $\tilde{e}_{\text{max}}$. To do this, recall the eigenvalue equations introduced in Section~\ref{subsection:QFT_LE}:
\begin{equation}
    \begin{dcases}
        \frac{1}{2^n}\sum_{y=0}^{2^j-1} \frac{\text{sin}\left(\pi(x-y)/2^j\right)}{\text{sin}\left(\pi(x-y)/2^n\right)} v_{n,j}^{k,y} = \lambda_{n,j}^{k} v_{n,j}^{k,x}\\
        \frac{1}{2^n}\sum_{y=0}^{2^{n-j}-1} \frac{\text{sin}\left(\pi(x-y)/2^{n-j}\right)}{\text{sin}\left(\pi(x-y)/2^n\right)} v_{n,n-j}^{k,y} = \lambda_{n,j}^{k} v_{n,n-j}^{k,x}\\
    \end{dcases}
\end{equation}
where the first equation corresponds to the upper partition and left singular vector, and the second equation corresponds to the lower partition and the right singular vector. In Section~\ref{subsection:QFT_LE} we see the actual left and right singular vectors are differed by a phase, i.e. the vectors $u_{n,j}^{k,x} = \omega_{n,j}^{x} v_{n,j}^{k,x}$ and $u_{n,n-j}^{k,x} = \omega_{n,n-j}^{x} v_{n,n-j}^{k,x}$. In addition, the left singular vectors $u_{n,j}^{k,x}$ have bits represented in the reverse order, i.e. $R_{j}|u_{n,j}^{k}\rangle$. This suggests the operator Schmidt decomposition of $\Omega_{n,j}$ is
\begin{equation}
    \Omega_{n,j} = \sqrt{2^n} \sum_{k=0}^{\chi - 1} \sigma_{n,j}^{k} \text{diag}(R_j|u_{n,j}^{k}\rangle) \otimes \text{diag}(|u_{n,n-j}^{k}\rangle)
\end{equation}
where $\text{diag}(|v\rangle)$ denotes the diagonal matrix with diagonal elements generated from the vector $|v\rangle$. The error can be then calculated as follows:
\begin{equation}
    \begin{split}
        &\tilde{e}_{\text{max}}
        \\ = & \sup_{|\phi\rangle} \left\|\sqrt{2^n}\sum_{k \ge \chi}\sigma_{n,j}^{k} ( A_{n,j}^{k}  \otimes  B_{n,j}^{k}  ) |\phi\rangle \right\|^2 \\
         =& 2^n \left\|\sum_{k \ge \chi}\sigma_{n,j}^{k} \text{diag}(R_{j}|u_{n,j}^{k}\rangle) \otimes \text{diag}(|u_{n,n-j}^{k}\rangle) \right\|^2 \\
        \leq& 2^n \left( \sum_{k \ge \chi}\sigma_{n,j}^{k} \max_{x}(|\omega_{n,j}^{x} v_{n,j}^{k,x}|) \max_{x}(|\omega_{n,n-j}^{x} v_{n,n-j}^{k,x}|) \right)^2 \\
        \leq& 2^n \left( \sum_{k \ge \chi}\tilde{\sigma}_{j}^{k} \max_{x}(|v_{n,j}^{k,x}|) \max_{x}(|v_{n,n-j}^{k,x}|) \right)^2 \\
    \end{split}
    \label{eq:e_max_app}
\end{equation}
where we can see the extra phases and reordering are also irrelevant to the error bounds. We now consider the behavior of the error in the limit $n \rightarrow \infty$, where $|v_{n,j}^{k}\rangle$ approaches the DPSSs $|\tilde{v}_{j}^{k}\rangle$
\begin{equation}
    \sum_{y=0}^{2^j-1} \frac{\text{sin}(\pi(x-y)/2^j)}{\pi(x-y)} \tilde{v}_{j}^{k,y} = \tilde{\lambda}_{j}^{k} \tilde{v}_{j}^{k,y}
    \label{eq:DPSS_app}
\end{equation}
while $|v_{n,n-j}^{k}\rangle$ approaches some continuous functions by the principle of Riemann sum
\begin{equation}
    \int_{0}^{1/2^j} \frac{\text{sin}\left(2^j\pi(\tilde{x}-\tilde{y})\right)}{\sin (\pi(\tilde{x}-\tilde{y}))} f_{j}^{k}(\tilde{y}) ~d\tilde{y} = \tilde{\lambda}_{j}^{k} f_{j}^{k}(\tilde{x})
\end{equation}
where $\tilde{x}, \tilde{y}$ are the continuous representation of $x/2^n, y/2^n$, $d\tilde{y}$ is $1/2^n$ in the limit, and the continous function $f_{j}^{k}(\tilde{x})$ is $\sqrt{2^n}v_{n,n-j}^{k,x}$ in the limit. The extra $\sqrt{2^n}$ in $f_{j}^{k}$ ensures it has a non-vanishing inner product with itself over range $[0,1/2^j]$, i.e.
\begin{equation}
\begin{split}
    &\int_{0}^{1/2^j} \left(f_{j}^{k}(\tilde{x})\right)^* f_{j}^{k}(\tilde{x}) ~ d\tilde{x} \\
    = & \lim_{n\rightarrow \infty} \sum_{x=0}^{2^{n-j}-1} \sqrt{2^n}\left(v_{n,n-j}^{k,x}\right)^* \sqrt{2^n}v_{n,n-j}^{k,x} \frac{1}{2^n}\\
    = & \lim_{n\rightarrow \infty} \sum_{x=0}^{2^{n-j}-1} \left(v_{n,n-j}^{k,x}\right)^* v_{n,n-j}^{k,x} \\
    = &1
\end{split}
\end{equation}
and thus $f_{j}^{k}$ has a well-defined value on every point in the region. In fact, we have seen in Appendix~\ref{appendix:spectral_concentration} that $f_{j}^{k}$ belong to a special kind of functions known as \emph{discrete prolate spheroidal wave functions} (DPSWFs), with coordinates shifted by $W=1/2^{j+1}$. DPSWFs are known to have the same eigenvalues as their corresponding DPSSs \cite{Slepian_1978}, and this property is illustrated here as $|\tilde{v}_{j}^{k}\rangle$ and $f_{j}^{k}$ arise from the same partition $j$. Therefore, as $n$ increases the maximum of $|v_{n,j}^{k,x}|$ and $\sqrt{2^n}|v_{n,n-j}^{k,x}|$ will approach the maximum of their limiting cases, which will be only dependent on $j$:
\begin{equation}
    \begin{dcases}
    \max_{x}(|v_{n,j}^{k,x}|) \rightarrow \max_{x}(|\tilde{v}_{j}^{k,x}|) \\
    \sqrt{2^{n}} \max_{x}(|v_{n,n-j}^{k,x}|) \rightarrow \max_{\tilde{x}}(|f_{j}^{k}(\tilde{x})|)
    \end{dcases}
\end{equation}
Therefore, the upper bound on the maximal error for truncation is $n$-independent, i.e.
\begin{equation}
    e_{\max}(D_{n,j}^{\chi}) \leq \left( \sum_{k \ge \chi}\tilde{\sigma}_{j}^{k} \max_{x}(|\tilde{v}_{j}^{k,x}|) \max_{\tilde{x}}(|f_{j}^{k}(\tilde{x})|) \right)^2
    \label{eq:n-independent_emax}
\end{equation}
We now use induction to show that once the upper bound is $n$-independent, it is also $j$-independent. First, we note that $Q_n$ is symmetrical under bit-reversal followed by transpose, i.e. $(R_{n}^\dagger Q_n R_{n})^T = Q_n$, which means we can get the exact same argument by replacing $j$ with $n-j$. This means $\tilde{e}_{\max}$ on $D_{n,j}^{\chi}$ share the same bound both with those on $D_{n+1,j}^{\chi}$ and $D_{n+1,j+1}^{\chi}$, therefore  $\tilde{e}_{\max}$ on $D_{n+1,j}^{\chi}$ and $D_{n+1,j+1}^{\chi}$ will also share the same bound, which makes it $j$-independent. For details see the diagrammatical illustration in Fig.~\ref{fig:QFT_error_relation}. Therefore, due to the decay of $\tilde{\sigma}_{j}^{k}$ the maximal error is also approximately $O(e^{-\chi \log(\chi/3)}/\sqrt{\chi})$, and thus at most $O(n e^{-\chi \log(\chi/3)}/\sqrt{\chi})$ error among the entire QFT-MPO.

\section{Functions used for timing of superfast Fourier transform}
\label{appendix:functions}

For the timing of the superfast Fourier transform algorithm based on the MPO representation of the QFT, we used three example functions, all defined on the domain $0\leq x \leq 1$.

The first function labeled \emph{1 Cosine} in Fig.~\ref{fig:qft_timings} is given by
\begin{equation}
    f_1(x) = \cos{(2\pi x)}
\end{equation}
It compresses exactly into an MPS with uniform bond dimensions all of the size $\chi_m=2$.

The second function labeled \emph{20 Cosines} is given by
\begin{equation}
    f_2(x) = \sum_{j=1}^{20} \cos{\big[1.1 \cdot (4j-2)\cdot(x-\frac{1}{2}) \big]}
\end{equation}
When representing this function as an MPS with $n=12$, for example, the bond dimensions were $[2, 4, 6, 7, 8, 9, 11, 13, 8, 4, 2]$.

The second function labeled \emph{1 Cosine + cusps} is given by
\begin{align}
    f_3(x) & =  \cos{(2\pi x)} + 2 e^{-3|x-0.4|} + e^{-2|x-0.7|} \nonumber \\
    &  + 2 e^{-3|x-0.8|} + e^{-2|x-0.9|}
\end{align}
When representing this function as an MPS with $n=12$, for example, the bond dimensions were $[2, 4, 7, 8, 8, 9, 10, 10, 8, 4, 2]$.

\end{appendices}

\clearpage

\nocite{*}
\bibliographystyle{apsrev4-2}
\bibliography{references}

\end{document}